\documentclass[12pt,preprint]{emulateapj}
\shorttitle{Optically Faint Variable Object Survey in the SXDF
}
\shortauthors{T. Morokuma et al.}
\begin{document}
\title{
The Subaru/XMM-Newton Deep Survey (SXDS) - V. Optically Faint Variable Object Survey
\footnote{Based in part on data collected at Subaru Telescope, 
which is operated by the National Astronomical Observatory of Japan. 
Based on observations (program GN-2002B-Q-30) obtained at 
the Gemini Observatory, which is operated by the Association of 
Universities for Research in Astronomy, Inc., under a cooperative 
agreement with the NSF on behalf of the Gemini partnership: 
the National Science Foundation (USA), the Particle Physics and
Astronomy Research Council (UK), the National Research Council
(Canada), CONICYT (Chile), the Australian Research Council
(Australia), CNPq (Brazil) and CONICET (Argentina).
}
}
\author{
Tomoki Morokuma\altaffilmark{1,2}, 
Mamoru Doi\altaffilmark{2}, 
Naoki Yasuda\altaffilmark{3}, 
Masayuki Akiyama\altaffilmark{4}, 
Kazuhiro Sekiguchi\altaffilmark{1,4}, 
Hisanori Furusawa\altaffilmark{4}, 
Yoshihiro Ueda\altaffilmark{5}, 
Tomonori Totani\altaffilmark{5}, 
Takeshi Oda\altaffilmark{5,6}, 
Tohru Nagao\altaffilmark{1,6}, 
Nobunari Kashikawa\altaffilmark{1}, 
Takashi Murayama\altaffilmark{7}, 
Masami Ouchi\altaffilmark{8,9}, 
Mike G. Watson\altaffilmark{10}, 
Michael W. Richmond\altaffilmark{11}, 
Christopher Lidman\altaffilmark{12}, 
Saul Perlmutter\altaffilmark{13}, 
Anthony L. Spadafora\altaffilmark{13}, 
Greg Aldering\altaffilmark{13}, 
Lifan Wang\altaffilmark{13,14}, 
Isobel M. Hook\altaffilmark{15}, 
Rob A. Knop\altaffilmark{16}
}
\altaffiltext{1}{Optical and Infrared Astronomy Division, National Astronomical Observatory, 2-21-1 Osawa, Mitaka, Tokyo 181-8588, Japan}
\altaffiltext{2}{Institude of Astronomy, Graduate School of Science, University of Tokyo, 2-21-1, Osawa, Mitaka, Tokyo 181-0015, Japan}
\altaffiltext{3}{Institute for Cosmic Ray Research, University of Tokyo, Kashiwa, Chiba 277-8582, Japan}
\altaffiltext{4}{Subaru Telescope, National Astronomical Observatory of Japan, 650 North A'ohoku Place, Hilo, HI 96720, USA}
\altaffiltext{5}{Department of Astronomy, Kyoto University, Sakyo-ku, Kyoto 606-8502, Japan}
\altaffiltext{6}{JSPS Fellow}
\altaffiltext{7}{Astronomical Institute, Graduate School of Science, Tohoku University, Aoba, Sendai 980-8578, Japan}
\altaffiltext{8}{Space Telescope Science Institute, 3700 San Martin Drive, Baltimore, MD 21218, USA}
\altaffiltext{9}{Hubble Fellow}
\altaffiltext{10}{Department of Physics and Astronomy, University of Leicester, Leicester LE1 7RH, UK}
\altaffiltext{11}{Physics Department, Rochester Institute of Technology, Rochester, NY 14623, USA}
\altaffiltext{12}{European Southern Observatory, Alonso de Cordova 3107, Vitacura, Casilla 19001, Santiago 19, Chile}
\altaffiltext{13}{Lawrence Berkeley National Laboratory, 1 Cyclotron Road, Berkeley, CA 94720, USA}
\altaffiltext{14}{Physics Department, Texas A\&M University, College Station, TX 77843, USA}
\altaffiltext{15}{University of Oxford Astrophysics, Denys Wilkinson Building, Keble Road, Oxford OX1 3RH, UK}
\altaffiltext{16}{Department of Physics and Astronomy, Vanderbilt University, P.O. Box 1807, Nashville, TN 37240. USA}
\email{tmorokuma@optik.mtk.nao.ac.jp}

\begin{abstract}
We present our survey for optically faint variable
objects using multi-epoch ($8-10$ epochs over $2-4$ years)
$i'$-band imaging data obtained with Subaru Suprime-Cam over $0.918$ deg$^2$
in the Subaru/XMM-Newton Deep Field (SXDF).
We found $1040$ optically variable objects by image subtraction for all
the combinations of images at different epochs.
This is the first statistical sample of variable objects at depths
achieved with 8-10m class telescopes or HST.                                  
The detection limit for variable components is $i'_{\rm{vari}}\sim25.5$ mag.
These variable objects were classified into variable stars, supernovae (SNe),
and active galactic nuclei (AGN),
based on the optical morphologies, magnitudes, colors, and optical-mid-infrared colors
of the host objects, spatial offsets of variable components from the host objects, and light curves.
Detection completeness was examined by simulating light curves
for periodic and irregular variability.                
We detected optical variability for $36\pm2\%$
($51\pm3\%$ for a bright sample with $i'<24.4$ mag)
of X-ray sources in the field.
Number densities of variable obejcts as functions of time intervals $\Delta{t}$ 
and variable component magnitudes $i'_{\rm{vari}}$ are obtained.
Number densities of variable stars, SNe, and AGN are
$120$, $489$, and $579$ objects deg$^{-2}$, respectively.
Bimodal distributions of variable stars in the color-magnitude diagrams indicate that
the variable star sample consists of bright ($V\sim22$ mag) blue variable stars of the halo population
and faint ($V\sim23.5$ mag) red variable stars of the disk population.
There are a few candidates of RR Lyrae providing
a possible number density of $\sim10^{-2}$ kpc$^{-3}$
at a distance of $>150$ kpc from the Galactic center.
\end{abstract}

\section{Introduction}\label{sec:intro}
It is well-known that there are a wide variety of objects showing optical variability 
by various mechanisms in various time scales in the universe. 
For example, stars in the instability strip in a color-magnitude diagram show periodic variability 
by pulsations (Cepheids and RR Lyrae), 
while cataclysmic variables show emergent variability induced by accretion flows 
from companion stars (novae), magnetic reconnection or rotations (dwarf stars). 
Thermonuclear or core-collapse runaway explosions of massive stars are 
observed as supernovae (SNe). 
Recent studies on optical afterglows of gamma-ray bursts (GRBs), which are extremely 
energetic phenomena, are starting to reveal their nature. 
Active galactic nuclei (AGN) show irregular flux variability not only in the optical but also in other wavelengths. 
Optically variable objects described above change their intrinsic brightness, 
but observed variability can be caused by other reasons. 
Searches for microlensing events have aimed to test the hypothesis that a significant fraction of the dark matter 
in the halo of our Galaxy could be made up of MAssive Compact Halo Objects (MACHOs) such as 
brown dwarfs or planets. 
Many asteroids in solar system have been found as moving objects. 
Many surveys for these optically variable objects have been carried out and have 
contributed to many important astronomical topics. 

The {\it cosmic distance ladder}, for example, were established partly based on studies of 
optically variable objects which are {\it standardizable} candles. 
Observations of variable stars in nearby galaxies enable us to determine distances to them by 
period-luminosity relations of Cepheids \citep[][and references therein]{sandage2006}. 
Recent deep high-resolution monitoring observations with Hubble Space Telescope (HST) increased 
the number of nearby galaxies whose distances were measured using Cepheids. 
Well-sampled light curves of type Ia supernovae (SNe Ia) in galaxies with distance measurements 
by Cepheids \citep{saha2006} showed empirical relationships between their light curve shapes 
and their luminosity: intrinsically brighter SNe Ia declined more slowly in brightness \citep{phillips1993}. 
These relations led to the discoveries of an accelerated expansion of the universe by 
two independent SN Ia search teams \citep*{perlmutter1998,riess1998}. 
In ongoing and planned SN surveys, SNe Ia are expected to set tight constraints 
on the value of the dark energy and its cosmological evolution if any. 

Variable stars have been also used to trace Galactic structure as standard candles. 
RR Lyrae are old ($>8$ Gyr) bright ($M_V\sim0.6$ mag) low-mass pulsating stars with blue colors of $B-V\sim0.3$ 
and are appropriate for tracing structures in the halo. 
They can be selected as stars showing large variability ($0.5-1.0$ mag in $V$-band) 
in short periods of $0.3-0.5$ days \citep{vivas2004}. 
RR Lyrae selected using multi-epoch imaging data obtained by the Sloan Digital Sky 
Survey \citep[SDSS;][]{york2000} provided a possible cut-off of the Galactic halo 
at $\sim65$ kpc from the Galactic center \citep{ivezic2000}. 
The subsequent studies on RR Lyrae by SDSS \citep*{ivezic2004a,ivezic2005,sesar2007} 
and the QUasar Equatorical Survey Team \citep[QUEST;][]{vivas2004,vivas2006} 
found many substructures of the Galactic halo including those already known. 

Optical variability of the first luminous AGN (quasar) 3C 273 were recognized \citep{smith1963} 
just after its discovery \citep{schmidt1963}. 
Since then, it has been known that almost all quasars show optical variability; 
indeed, it is one of common characteristics of AGN. 
Quasar surveys such as SDSS and the 2-degree Field Quasar Redshift Survey \citep[2QZ;][]{boyle2000} 
have been carried out mainly using optical multi-color selections. 
These surveys have made large catalogs of quasars 
\citep*{croom2004,schneider2005} and also found very high-$z$ quasars 
close to the reionization epoch \citep{fan2006a,fan2006b}. 
But efficiency of finding lower-luminosity AGN using color selections 
are expected to be low because AGN components get fainter compared with their host galaxy components. 
On the other hand, recent deep X-ray observations with ASCA, XMM-Newton, and Chandra satellites effectively 
found many distant low-luminosity AGN as well as obscured AGN. 
This high efficiency owes to faintness of host galaxies in X-ray and high transparency 
of dust to X-ray photons. 
X-ray observations can identify faint AGN easily while finding even unobscured AGN is 
difficult using optical color selections. 
However, since deep X-ray surveys over wide fields require 
a lot of telescope time \citep[e.g. $2$ Ms exposure in the Chandra Deep Field-North;][]{brandt2001}, 
optical variability is being recognized again as a good tracer for AGN. 
Some studies have succeeded in detecting optical variability of quasars \citep*{hook1994,giveon1999,hawkins2002}, 
and finding many quasars by optical variability with the completeness as high as classical 
UV excess selections \citep*{hawkins1993,ivezic2003}. 
Comparison of SDSS imaging data with older plate imaging data \citep*{devries2003,devries2005,sesar2006} 
and SDSS spectrophotometric data \citep{vandenberk2004} showed a clear anti-correlation between 
quasar luminosity and optical variability amplitude, as indicated in previous studies \citep*{hook1994,giveon1999}. 
Therefore, optical variability can be an efficient tool to find low-luminosity AGN 
if variable components can be extracted. 
Ultra-deep optical variability surveys with Wide-Field Planetary Camera 2 (WFPC2) 
installed on HST actually found 24 galaxies with variable nuclei down to $V_{\rm{nuc}}=27.5$ mag 
and $I_{\rm{nuc}}=27$ mag, which are as faint as nearby Seyfert 
galaxies ($-15<M_B<-19$) at $z\sim1$ \citep*{sarajedini2000,sarajedini2003}. 
\citet{cohen2006} also found several tens of AGN by optical variability 
using multi-epoch data with Advanced Camera for Survey (ACS) on board HST. 
The number densities of AGN selected via optical variability can be of the same order 
as those selected via deep X-ray observations \citep{brandt2005}. 
However, samples from the HST imaging data were not large enough for statistical studies. 
The comparable number densities of variability-selected AGN may indicate that there could be 
several populations of AGN with different properties. 
Furthremore, a search for faint transient objects with Suprime-Cam \citep{miyazaki2002} 
on Subaru telescope revealed very faint 
AGN variability in the nuclei of apparently normal galaxies, 
and found that such nuclei show very rapid ($\sim$ a few days) nuclear variability 
with a large fractional amplitude \citep{totani2005}. 
Such behavior is similar to that of the Galactic center black hole, Sgr A$^\ast$, 
rather than bright AGN, possibly indicating different physical nature of 
accretion disks of very low luminosity AGN and very luminous AGN. 

Object variability could affect searches for 
rare objects using non-simultaneous observational data. 
\citet{iye2006} and \citet{ota2007} searched for Lyman-$\alpha$ emitters 
(LAEs) at $z\sim7.0$ by comparing narrow-band data 
with broad-band data obtained more than a few years earlier. 
They succeeded in identifying a bright candidate spectroscopically as a LAE at $z=6.94$. 
Another candidate without spectroscopic identifications could be just a transient object 
and the authors treated it as a marginal candidate. 
SN rate studies based on optical variability 
\citep*{pain2002,dahlen2004,barris2006,sullivan2006,poznanski2007,oda2007} 
have sometimes confronted problems in classifying events due to 
insufficient observational data such as 
spectroscopic identifications, sufficient time samplings, and multi-wavelength measurements. 
Given limited time on large telescopes, one cannot make spectroscopic observations 
of all variable candidates; thus, 
multi-wavelength imaging data and light curves in year-scale baselines are 
very useful to separate SNe from other kinds of variable objects. 

Recently, many optical variability surveys with dense time samplings 
have been conducted for various astronomical purposes. 
Dividing exposure time into multiple epochs in extremely deep surveys using large or space 
telescopes also enables us to explore optical faint variability. 
Projects conducted recently or ongoing are 
Faint Sky Variability Survey \citep[FSVS;][]{groot2003}, 
Deep Lens Survey \citep[DLS;][]{becker2004}, 
Supernova Legacy Survey \citep[SNLS;][]{astier2006}, 
Great Observatory Origins Deep Survey \citep[GOODS;][]{strolger2004}, 
Groth-Westphal Survey Strip \citep[GSS;][]{sarajedini2006}, 
Hubble Deep Field-North \citep*[HDF-N;][]{sarajedini2000,sarajedini2003}, 
Hubble Ultra Deep Field \citep[HUDF;][]{cohen2006}, 
Sloan Digital Sky Survey-II (SDSS-II) Supernova Survey 
\citep{sako2007}. 
Densely sampled observations targeting GRB orphan afterglows 
also have been carried out \citep[e.g.,][]{rau2006}. 
We focus here on our survey for optically faint variable objects since 2002 
in the Subaru/XMM-Newton Deep Field \citep[SXDF,][]{sekiguchi2004,sekiguchi2007}. 
The data have been taken by the Subaru/XMM-Newton Deep Survey (SXDS) project. 
Our study described in this paper provides the first statistical sample 
of optically faint variable objects and is unique among studies using 8-10m class 
telescopes and HST in its combination of wide field coverage and depth. 
The Suprime-Cam imaging data and the analysis for finding optically variable objects are described 
in \S\ref{sec:ofvos}, followed by other observational data in \S\ref{sec:addobs}. 
We describe object classifications in \S\ref{sec:objclass} 
and detection completeness in \S\ref{sec:completeness}. 
The results are discussed for statistics of the whole sample in \S\ref{sec:overall_nc} 
and for variable stars alone in \S\ref{sec:varistar}. 
We summarize our results in \S\ref{sec:summary}. 
In this paper, we adopt throughout the AB magnitude system for optical and mid-infrared photometry. 

\section{Optically Faint Variable Object Survey with Subaru Suprime-Cam}\label{sec:ofvos}
\subsection{Subaru/XMM-Newton Deep Field (SXDF)}\label{sec:sxdf}
We are carrying out an optically faint variable object survey using a wide-field camera, Suprime-Cam, 
on the prime focus of Subaru 8.2-m telescope. 
The widest field-of-view ($34'\times27'$) among the optical imaging 
instruments installed on HST and 8-10m telescopes provides us a unique 
opportunity for statistical studies of rare and faint objects which are 
less affected by cosmic variance than those derived from smaller fields. 

The SXDF is 
centered on (02$^{\rm{h}}$18$^{\rm{m}}$00$^{\rm{s}}$, -05:00:00) in J2000.0, 
in the direction towards the Galactic halo, $(l,b)=(169^\circ,-60^\circ)$. 
The SXDF is an on-going multi-wavelength project from X-ray to radio \citep*{sekiguchi2004,sekiguchi2007}, 
exploring aspects of the distant universe 
such as the nature of the extragalactic X-ray populations \citep*{watson2005,ueda2007,akiyama2007}, 
large-scale structures at high redshift \citep*{ouchi2005a,ouchi2005b} and 
cosmic history of mass assembly \citep*{kodama2004,yamada2005,simpson2006b}. 
The field coverage is $\sim1.2$ deg$^2$, which consists of five pointings of Suprime-Cam 
(SXDF-C, SXDF-N, SXDF-S, SXDF-E, and SXDF-W, respectively, see Figure \ref{fig:sxdfimage2}). 
We use the $i'$-band-selected photometric catalogs 
in $B$, $V$, $R_C$, $i'$ and $z'$-bands, 
which are summarized in \citet{furusawa2007}. 
The depths of the catalogs which we use are almost the same among the five fields, 
$28.4$, $27.8$, $27.7$, $27.7$, and $26.6$ mag ($3\sigma, 2\farcs0\phi$) 
in $B$, $V$, $R_C$, $i'$ and $z'$-bands, respectively (Table \ref{tab:surveyarea}). 
Optical photometric information used in this paper except for variability measurements is derived from these catalogs 
and are not corrected for the Galactic extinction in the SXDF direction, 
$E(B-V)=0.019-0.023\ \rm{mag}$ \citep[][see \citealt{furusawa2007}]{schlegel1998}. 
\subsection{Subaru Suprime-Cam Imaging Data}\label{sec:spcamdata}
Our survey for optically faint variable objects in the SXDF is based on multi-epoch 
$i'$-band Suprime-Cam imaging data, starting in September 2002 \citep*{furusawa2007,yasuda2007}. 
The Suprime-Cam observations are summarized in Table \ref{tab:spcamobs}. 
These observations include those for an extensive high-$z$ SN search using well-sampled $i'$-band 
images in 2002 \citep*{yasuda2003,yasuda2007} with which \citet{doi2003} reported 
the discoveries of $13$ high-$z$ SNe. 
In addition to observations used to make the catalogs in \citet{furusawa2007}, 
we carried out $i'$-band observations in 2005. 
The final images were taken in October 2003 in the two fields (SXDF-N and SXDF-W) 
and in September 2005 in the remaining three fields (SXDF-C, SXDF-N, and SXDF-E). 
The numbers of the observational epochs ($N_{\rm{epoch}}$) 
in the five fields are $10$, $8$, $8$, $10$, and $8$, respectively. 
Time intervals of the observations are from $1$ day to $3$ years in observed frame. 

In each epoch, stacked images were made in a standard method for the Suprime-Cam data 
using the NEKO software \citep{yagi2002} and the SDFRED package \citep{ouchi2004}. 
Some images taken on different dates 
were combined together to make all the depths almost the same. 
Then, we geometrically transformed the images using {\it geomap} and {\it geotran} tasks 
in IRAF\footnote{IRAF is distributed by the National Optical Astronomy Observatories, 
which are operated by the Association of Universities for Research in Astronomy, 
Inc., under cooperative agreement with the National Science Foundation.} 
in order to match the coordinates in better accuracy each other. 
The root-mean-square of residuals of the coordinate differences are typically $0.4$ pixel. 

Typical exposure times of one hour provide limiting magnitudes of 
$m_{\rm{lim}}=25.2-26.8$ mag ($<m_{\rm{lim}}>_{\rm{median}}=26.0$ mag, $5\sigma, 2\farcs0\phi$). 
The full-width-at-half-maxima (FWHM) of point spread function (PSF) were 
$\theta=0\farcs52$-$1\farcs54$ ($<\theta>_{\rm{median}}=0\farcs70$, see Table \ref{tab:spcamobs})
\footnote{Pixel scale of Suprime-Cam is $0\farcs202$ pixel$^{-1}$.}. 

\begin{figure}[htbp]
\begin{center}
\epsscale{.8}
\includegraphics[angle=270,scale=0.40]{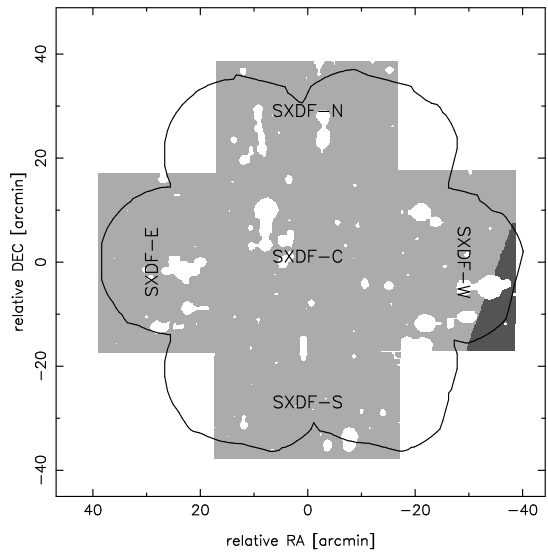}
\caption{The Suprime-Cam imaging fields (light gray and dark gray regions) in the SXDF, 
superimposed XMM-Newton EPIC imaging fields (black solid line) and Spitzer IRAC imaging fields 
(light gray region). 
The coordinates are measured relative to the center of the SXDF, 
($02^{\rm{h}}18^{\rm{m}}00^{\rm{s}}$, -05:00:00) in J2000.0. 
Our optically variable object survey has been carried out in 
the entire Suprime-Cam fields over $0.918$ deg$^2$ except for regions around bright objects (white blanks). 
The five pointings of the Suprime-Cam, SXDF-C, SXDF-N, SXDF-S, SXDF-E, and SXDF-W, are described. 
\label{fig:sxdfimage2}}
\end{center}
\end{figure}

For our variability study, we used regions overlapping in all the epochs in each field. 
Considering saturation and non-linearity of CCDs at high signal levels, 
we excluded regions around bright objects to ensure reliable detection of object variability. 
The total effective area is $0.918$ deg$^2$ (Table \ref{tab:surveyarea}) shown as 
light gray and dark gray regions in Figure \ref{fig:sxdfimage2}. 

\subsection{Variability Detection}\label{sec:varidetect}
\begin{figure}[htbp]
\begin{center}
\epsscale{.8}
\includegraphics[angle=270,scale=0.40]{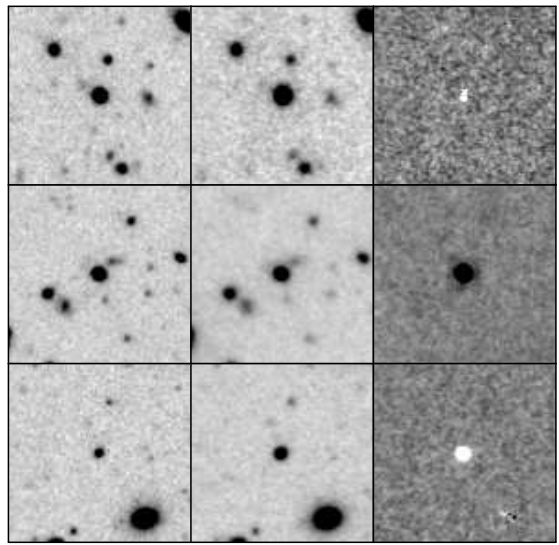}
\caption{Examples of variability detection. 
In each row, an image on a reference date, an image on another date, 
and subtracted image are shown from left to right. 
Variable objects are clearly seen as positive or negative residuals in the right panels. 
All these three objects were spectroscopically identified. 
The top object is a M dwarf star, the middle object is a broad-line AGN 
at $z=2.150$, and the bottom object is a SN Ia at $z=0.606$. 
The spectra and light curves are shown in the top row of Figure \ref{fig:specvaristar}, 
the third row of Figure \ref{fig:specagnvari}, 
and the bottom row of Figure \ref{fig:specsnvari}, respectively. 
The box sizes are $20$''. 
\label{fig:imsub}}
\end{center}
\end{figure}

The Suprime-Cam images have different PSF size (and shape) in every epoch 
because of time-varying atmospheric seeing. 
Therefore, we can not measure variability of objects by simply comparing flux 
within a fixed small aperture as \citet{sarajedini2000} and \citet{sarajedini2003} 
did for the HST WFPC2 images. 
\citet{cohen2006} used total magnitude for variability detections 
in the images obtained with the Advanced Camera for Survey (ACS) on HST 
because the PSF varied with locations on the CCDs of ACS. 
Accurate photometry itself of any kind for extended objects is often difficult. 
Since variable components are almost always point sources, the light from non-variable 
extended components just increases the noise for variability detections. 
Hence, we used an image subtraction method introduced by \citet{alard1998} 
and developed by \citet{alard2000} instead of using total magnitudes of objects. 
This subtraction method using space-varying convolution kernels, 
which are obtained by fitting kernel solutions in small sub-areas, 
enables us to match one image against another image with a different PSF. 
We can then detect and measure variable objects in the subtracted images. 
We applied this method for all the possible two pairs 
\footnote
{
The number of the combinations is $_{N_{\rm{epoch}}}C_2$.
}
of the stacked images at different epochs
and detected variability of the objects in each field. 
Figure \ref{fig:imsub} shows examples of the image subtration for three of the variable objects. 

We now describe details of the procedures for variability detection and 
our definitions of variable objects. 
First, we smoothed all the subtracted images with the PSFs and 
detected positive and negative peaks of surface brightness in subpixel unit. 
Then, in all the subtracted images, we selected objects whose aperture flux 
in a fixed diameter of 2\farcs0 were above $5\sigma_{\rm{bf}}$ 
or below $-5\sigma_{\rm{bf}}$ of 
the background fluctuations ($\sigma_{\rm{bf}}$) within the same aperture. 
The background fluctuations $\sigma_{\rm{bf}}$ is obtained by calculating 
the standard deviation of 2\farcs0 aperture flux in $\sim200$ places 
around each positive or negative peak in the subtracted images. 
We inspected these variable candidates visually to exclude candidates around mis-subtracted regions. 
Second, we defined the central coordinates of the variable object candidates as the flux-weighted 
averages of all the coordinates where significant variability were detected 
in the subtracted images in the first procedure. 
Third, we did 2\farcs0-diameter aperture photometry for all these variable object 
candidates in the subtracted images, the reference images of which were an image 
with the smallest PSF size in each 
field\footnote{They are images obtained on 02/09/29 and 02/09/30 for all the fields.}, 
and scaled the aperture measurements to the total flux 
assuming that the variable components were always point sources. 
The factors to scale 2\farcs0 aperture flux to total flux were calculated as ratios 
between 2\farcs0 aperture flux and total flux for point sources in each image 
(typically $\sim1.1-1.3$). 
We constructed total flux light curves of all the candidates and defined 
optically variable objects as objects showing more than $5\sigma_{\rm{lc}}$ 
($\sigma_{\rm{lc}}$: photometric errors in the light curves) 
variability in at least one combination of the epochs in the light curves. 
We note that some variable objects have unreliable photometric points in the light curves. 
In the subtraction method, very slight offsets (even $\sim0.05$ pixel) for bright objects between two images 
which are used for the subtraction could provide significant {\it dipole} subtractions. 
When an object has both a positive peak above $5\sigma_{\rm{bf}}$ and a negative peak below $-5\sigma_{\rm{bf}}$ 
locally (around $\sim10$ pixel square) in a certain subtracted image 
used to make the light curve, 
the photometric point of the object at that epoch are regarded as unreliable. 
These epochs were not used in evaluating variability. 

We followed these procedures for all the five fields. 
Finally, we made an optically variable object catalog consisting of $1184$ objects 
in the area of $0.918$ deg$^2$ ($1290$ variable objects deg$^{-2}$). 
However, for widely spread galaxies or bright objects, 
subtracted images are sometimes insecurely extended and show bumpy features 
although saturated or non-linearity pixels were removed from the detections in advance. 
Variability of these objects are considered marginal. 
When we exclude such marginally variable objects, the number of variable objects is $1040$ 
($1133$ variable objects deg$^{-2}$). 
These $1040$ variable objects have $9236$ photometric points in their light curves, 
and $359$ unreliablly measured photometric points ($3.9\%$) of $169$ objects were not used in evaluating variability. 
In the following discussions, we use this secure sample of $1040$ variable objects. 
We note that detection completeness is discussed in \S\ref{sec:completeness}. 

In principle, variability should be measured as changes of luminosity of objects. 
To do so, it is necessary to measure total flux of variable objects. 
However, this is possible only for point sources such as variable stars and quasars whose 
luminosity is much larger than that of their host galaxies. 
To measure variable 
AGN separately from their host galaxies is difficult 
using ground-based imaging data especially when the flux of variable object is faint 
compared to the flux of the host. 
We measure the flux of the variable component ($\Delta{f}\equiv f_1-f_2$) as describe above. 
In this paper, we use variable component magnitudes 
($i'_{\rm{vari}}\equiv-2.5\log(|\Delta{f}|)+m_{\rm{zero}}$) 
to describe object variability, 
and we adopt the total magnitudes which are measured in the SXDF catalogs 
as the magnitudes of host objects. 

\subsection{Assignment of Host Objects}\label{sec:assignhost}
In order to investigate how far the variable components are located from the centers 
of possible host objects, and their properties such as magnitudes and colors in the catalogs, 
we assigned host objects to all the variable objects. 
In general, AGN and SNe have their host galaxies while variable stars have no host objects. 
In this paper, we defined variable stars themselves as host objects 
to treat all the variable objects equally. 
We calculated projected distances of the variable components from the surrounding objects 
in the stacked images taken when the variable objects were in the faintest phases, 
and defined the nearest objects as the host objects. 
The positions of the host objects were measured as peaks of surface brightness 
of the images smoothed by the PSFs. 
In the faintest phases of variable objects which may disappear (i.e. SNe), 
host objects should be detected, but high-$z$ SNe which ocuur in diffuse galaxies 
seem hostless in some cases. 
Assigned host objects for such SNe may not be real host galaxies. 
It is also possible that transient objects on faint host objects may have no equivalent 
host objects in the SXDF catalogs because the SXDF $i'$-band images 
for the catalogs were made by stacking all the exposures except for those in 2005. 
Among $1040$ variable objects, the number of such objects, which do not have 
equivalent objects in the SXDF catalogs, is $8$. 
These possible transient objects may be hostless SNe and 
are treated as SNe if their light curves satisfy 
a criterion described in \S\ref{sec:criteriasnagn2}. 
The number of the possible transient objects is small 
and we do not exclude or discriminate them in most of 
the following discussions. 
In \S\ref{sec:overall_nc}, we discuss number densities of transient
objects, which are defined as objects not detected in their faint
phases; objects appear in some epochs at positions where there are no
objects detected in other epochs. 
In this paper, the number of detections for each transient object 
is not considered in the definition of transient objects. 

Information on host objects, such as total magnitudes and colors, is derived 
from integrated images which were obtained 
in various epochs in each band as shown in \citet{furusawa2007} 
and they are averaged over the observational time spans. 
Therefore, properties of objects are appropriate for statistical studies 
and could be inaccurate if we focus on a certain object. 

\section{Additional Observational Data}\label{sec:addobs}
The SXDF project has carried out multi-wavelength observations from X-ray to radio 
and follow-up optical spectroscopy. 
In this paper, optical spectroscopic results and X-ray imaging data 
are used to confirm the validity of our object classifications. 
X-ray data is also used for evaluating the completeness of AGN variability detection. 
We use mid-infrared imaging data 
to classify optically variable objects in \S\ref{sec:objclass}.  
In this section, we summarize these observations and cross-identifications of optically variable objects. 
The summary of the X-ray and mid-infrared imaging data is given in Table \ref{tab:surveyarea}. 

\subsection{Optical Spectroscopy}\label{sec:optspec}
Since 2002, the follow-up optical spectroscopic observations have been conducted 
with many telescopes and instruments 
\citep*{yasuda2003,yamada2005,lidman2005,watson2005,ouchi2005a,ouchi2005b,yasuda2007}. 
Instruments and telescopes used are 
the Two-Degree Field (2dF) on the Anglo-Australian Telescope (AAT), 
the Faint Object Camera And Spectrograph \citep[FOCAS;][]{kashikawa2002} on Subaru telescope, 
the VIsible MultiObject Spectrograph \citep[VIMOS;][]{lefevre2003}, and 
the FOcal Reducer/low dispersion Spectrograph 2 (FORS2) on Very Large Telescope (VLT), 
the Gemini Multi-Object Spectrograph \citep[GMOS;][]{hook2004} on Gemini-North telescope, 
and the Echellette Spectrograph and Imager \citep[ESI;][]{sheinis2002} on Keck-II telescope. 
All the observations except for the Keck-II ESI, which is a high-dispersion 
echellete spectrograph, 
have been carried out in relatively low spectral resolution mode ($R\sim 500$). 
Most of the spectroscopic observations were done in multi-object spectroscopy mode. 

\begin{figure}[htbp]
\begin{center}
\epsscale{.8}
\includegraphics[angle=270,scale=0.320]{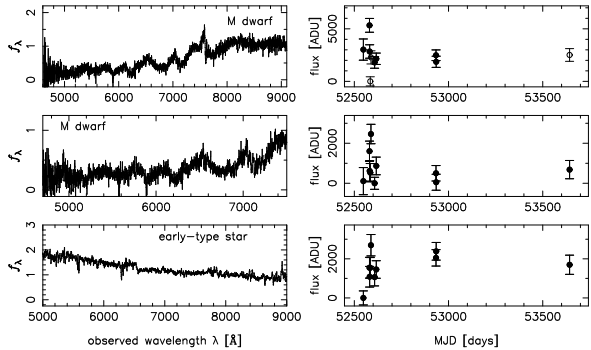}
\caption{Spectra in arbitrary unit and light curves of three variable stars. 
All the spectra were obtained with FOCAS on Subaru telescope. 
The observational configurations were the 300B grism and SY47 order-sort filter for the top two 
stars and the 150 grism and SY47 order-sort filter for the bottom star. 
The top two rows are M dwarf stars with $B-V=1.34$ and $1.36$. 
The bottom row is an early-type star with $B-V=0.44$. 
The light curves of differential flux were plotted in filled circles in linear scale (in ADU) 
because we can not plot magnitudes in their faintest phases. 
The zeropoint is $34.02$ mag. 
Unreliable photometric points are plotted in open circles. 
\label{fig:specvaristar}}
\end{center}
\end{figure}

\begin{figure}[htbp]
\begin{center}
\epsscale{.8}
\includegraphics[angle=270,scale=0.320]{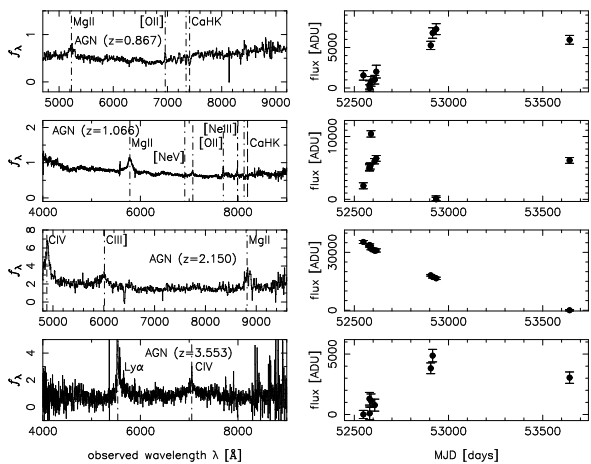}
\caption{Spectra in arbitrary unit and light curves of four extragalactic variable objects 
which are classified as AGN based on their light curves and variable locations ({\it central variability}). 
All the spectra were obtained with FOCAS on Subaru telescope. 
The observational configurations were the 300B grism without any order-sort filters for the top and third 
AGN and the 150 grism and SY47 order-sort filter for the second and bottom AGN. 
The redshifts are $0.867$, $1.066$, $2.150$, and $3.553$, from top to bottom. 
Dashed lines indicate detected emission and absorption lines. 
The zeropoint the light curves is $34.02$ mag. 
\label{fig:specagnvari}}
\end{center}
\end{figure}

\begin{figure}[htbp]
\begin{center}
\epsscale{.8}
\includegraphics[angle=270,scale=0.320]{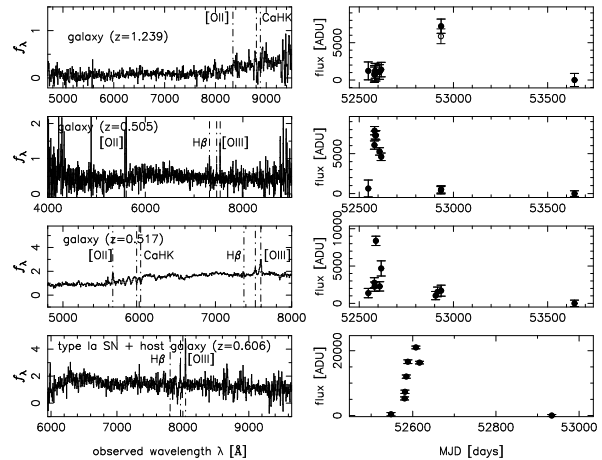}
\caption{Spectra in arbitrary unit and light curves of four extragalactic variable objects. 
The top three objects are classified as SNe based on their light curves and variable locations 
({\it offset variability}). 
The bottom object was spectroscopically identified as a SN Ia, SN 2002km \citep{lidman2005}, 
but we can not determine 
whether they are SNe or AGN because baselines of their light curves are not long enough 
(see \S\ref{sec:criteriasnagn2}). 
The top three objects were observed with FOCAS on Subaru telescope. 
The observational configurations were the 150 grism without any order-sort filters for the top 
and third objects, 
and the 300B grism and SY47 order-sort filter for the second object. 
The bottom object were observed with FORS2 on VLT 
using the 300I grism and the OG590 order-sort filter 
(see Figure A.32 in \citealt{lidman2005} for the SN component). 
The redshifts are $1.239$, $0.505$, $0.517$, and $0.606$, from top to bottom. 
Dashed lines indicate detected emission and absorption lines. 
The zeropoint is $34.02$ mag. 
\label{fig:specsnvari}}
\end{center}
\end{figure}

Out of $1040$ variable objects, $119$ objects were targeted for spectroscopy. 
Of these $119$ objects, $99$ objects were identified and the redshifts were determined. 
The spectroscopic sample includes $3$ stars, $14$ host galaxies of SNe, $6$ of which are 
spectroscopically identified as SNe Ia \citep*{lidman2005,yasuda2007}, 
$53$ broad-line AGN, $4$ \ion{[Ne}{5]} emitting galaxies and $25$ galaxies. 
Strong \ion{[Ne}{5]} emission lines are only observed among AGN, 
which emit strong ionizing photons above 97.2 eV, 
and \ion{[Ne}{5]} emitting galaxies are considered to host AGN. 
Other galaxies which appear to be normal galaxies also can host not only SNe but also AGN 
because many AGN in our sample are faint compared to the host galaxies 
so that the spectra may not show significant features of AGN origin.   
The redshifts of $96$ extragalactic variable objects are $0.240<z<4.467$. 
We show $11$ examples of the spectra obtained with FOCAS on Subaru and FORS2 on VLT 
in Figure \ref{fig:specvaristar}, Figure \ref{fig:specagnvari}, and Figure \ref{fig:specsnvari}. 
The top two rows of Figure \ref{fig:specvaristar} are those for M dwarf stars showing bursts in 2002 
with red optical colors of $B-V=1.34$ and $1.36$. 
The bottom row of Figure \ref{fig:specvaristar} is an early-type star with $B-V=0.44$. 
We also show spectra of four broad-line AGN at $z=0.867$, 
$1.066$, $2.150$, and $3.553$ in Figure \ref{fig:specagnvari}, 
and those of SN host galaxies at $z=1.239$, $0.505$, $0.517$, $0.606$ in Figure \ref{fig:specsnvari}. 
The light curves are also shown in the figures. 
The photometric points in these light curves are measured as differential flux in the subtracted images. 

Since spectroscopic observations of variable objects were mainly targeted 
for X-ray sources and high-$z$ SN candidates, the spectroscopic sample of 
variable objects are biased for those classes of variable objects. 
Therefore, we can not discuss much about fractions of variable objects 
only from the spectroscopic sample. 

\subsection{X-ray Imaging}\label{sec:xraydata}
X-ray imaging observations in the SXDF were carried out 
with the European Photon Imaging Camera (EPIC) on board XMM-Newton satellite. 
They consist of one deep ($\sim100$ ks) pointing on the center of the SXDF and six shallower ($\sim50$ ks) 
pointings at the surrounding regions. 
In total, X-ray imaging covers most of the Suprime-Cam imaging fields 
(see Figure \ref{fig:sxdfimage2}). 
The details of the XMM-Newton EPIC observations, data analyses, and optical 
identifications are described in \citet{ueda2007} and \citet{akiyama2007}. 
The limiting fluxes are down to $1\times10^{-15}$ erg$^{-1}$ cm$^{-2}$ s$^{-1}$ 
in the soft band (0.5-2.0 keV) and $3\times10^{-15}$ erg$^{-1}$ cm$^{-2}$ s$^{-1}$ 
in the hard band (2.0-10.0 keV), respectively. 

In this paper, we use a sample which was detected with likelihood 
larger than nine in either the soft or hard band and also within our variability survey fields. 
After excluding objects in the regions not used for variability detection (Table \ref{tab:surveyarea}), 
we have $481$ X-ray sources over $0.808$ deg$^2$ 
Among $936$ variable objects within the X-ray field, $172$ objects were detected in X-rays 
($165$ and $91$ objects detected in the soft and hard bands, respectively, see Table \ref{tab:surveyarea}). 

\subsection{Mid-Infrared Imaging}\label{sec:mirdata}
Mid-infrared imaging data in $3.6\mu$m, $4.5\mu$m, $5.8\mu$m, and $8.0\mu$m-bands with 
the InfraRed Array Camera \citep[IRAC;][]{fazio2004} on Spitzer space telescope 
were obtained in the SXDF as a part of the Spitzer Wide-area 
InfraRed Extragalactic \citep*[SWIRE;][]{lonsdale2003,lonsdale2004} survey. 
The IRAC data covers almost the entire Suprime-Cam field except for a part of SXDF-W. 
The covered field is $0.889$ deg$^2$ ($97\%$ of the Suprime-Cam field) shown as gray regions in Figure \ref{fig:sxdfimage2}. 
The reduced IRAC images were obtained from the SWIRE Archive and the catalogs were made by \citet{akiyama2007}. 
They used the IRAC imaging data for optical identifications of X-ray sources. 
In this paper, we use only the $3.6\mu$m-band data for the object classification in $\S$\ref{sec:objclass}. 
The limiting magnitude is $\sim22.0$ mag (total flux, $5\sigma$ for point sources). 
Among $1040$ variable objects, $1028$ objects are within the IRAC field 
and $995$ objects are detected in the $3.6\mu$m-band (Table \ref{tab:surveyarea}). 
Spatial resolution of the IRAC $3.6\mu$m-band is not high, $\sim1''$, 
but we can assign IRAC identifications to almost all the optically variable objects with good accuracy. 

\section{Object Classification}\label{sec:objclass}
\begin{figure}[htbp]
\begin{center}
\epsscale{.8}
\includegraphics[angle=270,scale=0.38]{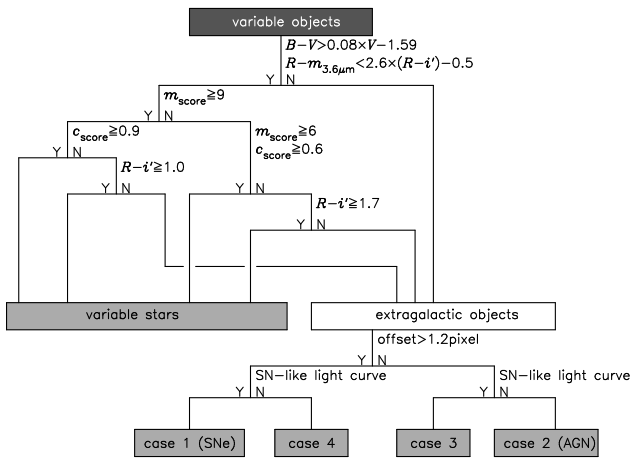}
\caption{Flow chart of object classification. 
See the text in \S\ref{sec:starselect} and \S\ref{sec:snagnseparate} for details. 
If an object satisfies a criterion at some point along the path, 
it goes to the left; if not, to the right. 
In the first criteria, the IRAC data is necessary, 
so that we can classify variable objects in $97\%$ of the overall field. 
The last criteria, using light curves for extragalactic objects, 
requires time baselines over longer than three years, 
and therefore only objects in the three fields, SXDF-C, SXDF-S, and SXDF-E can be classified. 
\label{fig:classificationflowchart}}
\end{center}
\end{figure}

The variable object sample includes several classes of variable objects. 
In this section, we classify these variable objects into variable stars, 
SNe, and AGN using optical and mid-infrared imaging parameters. 
The procedures described in \S\ref{sec:starselect} and \S\ref{sec:snagnseparate} 
are summarized in a flow chart of Figure \ref{fig:classificationflowchart}. 

We note again that the object photometric information is averaged 
over the observational time spans, but its effect on our statistical 
classifications and discussions is small even if objects are variable. 

\subsection{Star Selection}\label{sec:starselect}
We extracted variable stars from the variable object sample 
using the Suprime-Cam optical imaging and the IRAC mid-infrared imaging data. 
Since the IRAC data covers $97\%$ of the variability survey field in the SXDF, 
we concentrate on the $1028$ variable objects in the overlapped region of $0.889$ deg$^2$ 
for the following discussions such as object classifications and number densities. 
Star selection is based on the optical morphologies, optical magnitudes, optical colors, 
and optical and mid-infrared colors of the variable objects. 
We used total magnitudes in both optical and mid-infrared wavelengths to make this selection. 

\subsubsection{Criteria on Optical Morphology, Magnitude, and Color}\label{sec:criteriastar1}
First, we assigned to all the variable objects {\it morphological scores} ($m_{\rm score}=0-10$) 
and {\it color scores} ($c_{\rm score}=0.00-1.00$)
using a method described in \citet{richmond2005}. 
The {\it morphological scores} $m_{\rm score}$ were calculated based 
on magnitude differences $\delta\equiv m_2-m_3$ 
between 2\farcs0 aperture magnitudes ($m_2$) and 3\farcs0 aperture magnitudes ($m_3$), 
and the CLASS\_STAR values in all the five broad-bands from the SExtractor \citep{bartin1996} outputs 
in the SXDF catalogs. 
If an object has $0.10<\delta<0.20$, we add $1$ to $m_{\rm{score}}$. 
If a CLASS\_STAR value of an object is larger than some value 
(typically $0.90$, small variations from field to field)
we also add $1$ to $m_{\rm{score}}$. 
The {\it color scores} $c_{\rm score}$ were assigned based on the distances from the stellar loci 
in two color-color planes, $B-V$ versus $V-R$ and $V-R$ versus $R-i'$, using the SExtractor output IsophotMag. 
The stellar loci were defined using colors of a small set of bright stars. 
The color score represents the probability that each object is inside the stellar locus 
in color-color space using a Monte Carlo approach. 
A measure of the degree to which the colors of an object 
(including their uncertainties) fall within the stellar locus at the 
point of closest approach.  Values range from $0.0$ (color is completely
outside the stellar locus) to $1.0$ (color is completely within the locus).
Objects with larger values for both scores are more likely to be stars, 
and the most likely stars should have $m_{\rm score}\ge9$ and $c_{\rm score}\ge0.90$. 
However, faint, red or blue stars are likely to have lower values 
for both the scores because of the low signal-to-noise ratio of photometry in some bands. 
In this classification, stars redder than G stars were not used for determining the stellar loci. 
The reddest stellar locus is at $R-i'\sim1.7$. 
Therefore, red stars with $R-i'>1.7$ can not have high values of $c_{\rm score}$ 
and we do not use $c_{\rm{score}}$ values for red stars with $R-i'>1.7$. 
Galaxies with similar colors can contaminate into the sample. 
However, they can be excluded by an optical-mid-infrared color selection described below. 
Another problem is color differences of stars due to differences in their metallicity. 
Bright stars used for determining the stellar loci are mainly disk stars, 
while fainter stars are likely to belong to the halo population. 
Color differences in $R-i'$ caused by the population difference appear at $R-i'>1.0$, 
so, faint red stars with $R-i'>1.0$ tend to have low $c_{\rm{score}}$ values; 
we decided to classify point-like objects with $R-i'>1.0$ as stars. 
Hence, we first assigned stellar likelihood to all the variable objects 
through five criteria using only optical photometric information, 
\begin{itemize}
  \item[a)] stars with $m_{\rm score}\ge9$ and $c_{\rm score}\ge0.90$ (point-like objects with stellar colors), 
  \item[b-1)] probable stars with $9>m_{\rm score}\ge6$ and $0.90>c_{\rm score}\ge0.60$ 
(less point-like objects with less stellar colors), 
  \item[b-2)] probable stars with $m_{\rm score}\ge9$ and $R-i'>1.0$ (point-like objects with red colors), 
  \item[c)] possible stars with $R-i'>1.7$ (objects with very red colors), 
  \item[d)] non-stellar objects which do not satisfy any criteria above. 
\end{itemize}
Objects satisfying one of the criteria, a), b-1), b-2), or c) are considered as stars. 

In order to check the validity of our star selections based on only optical imaging parameters, 
we used a population synthesis model \citep[Besan\c{c}on model;][]{robin2003}. 
We investigated the expected distributions of stars including non-variable stars in the SXDF 
which were selected in the same criteria 
in a $B-V$ versus $V$ color-magnitude diagram. 
The distributions contain three sequences and two of them, sequences for younger 
disk population and older halo population are reasonably duplicated by the model.  
Another sequence of faint blue point-like objects has 
almost the same distribution as more extended galaxies in this diagram. 
Then, we concluded that these faint blue point-like objects are galaxies, 
and adopted one more criterion, $B-V>0.08\times V-1.59$ 
(dashed line in the top panel of Figure \ref{fig:colorcolor}), to exclude them 

\subsubsection{Criteria on Optical-Mid-Infrared Color}\label{sec:criteriastar2}
\begin{figure}[htbp]
\begin{center}
\epsscale{.8}
\includegraphics[angle=270,scale=0.30]{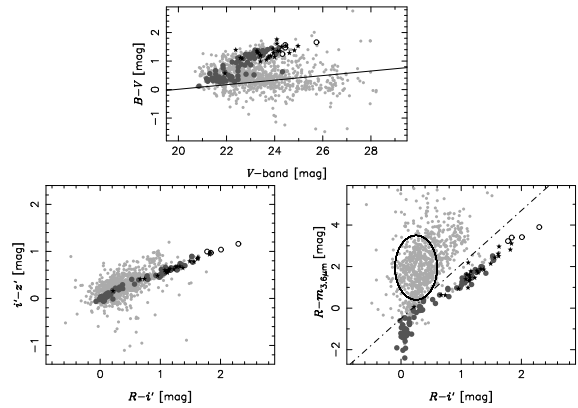}
\caption{
Color-color diagrams of variable objects in $R-i'$ versus $i'-z'$ (left panel) 
and $R-i'$ versus $R-m_{3.6\mu\rm{m}}$ (right panel). 
We plot all the variable objects in small gray triangles, 
{\it probable stars} in large open circles, and {\it reliable stars} in large filled circles. 
A criterion indicated in a dashed line in the top panel, $B-V>0.08\times V-1.59$, 
is adopted to exclude point-like blue galaxies. 
Dot-dashed line in the bottom right panel is a line separating stars and galaxies, 
$R-m_{3.6\mu\rm{m}}=2.6\times(R-i')-0.5$. 
A large ellipse indicates a region of SDSS quasar ($0.1<z<5.2$) 
colors, $r-i:r-m_{3.6\mu\rm{m}}$, 
in \citet{hatziminaoglou2005} and \citet{richards2006}. 
\label{fig:colorcolor}}
\end{center}
\end{figure}

Second, we investigated optical-mid-infrared colors of the variable objects 
in a $R-i'$ versus $R-m_{3.6\mu\rm{m}}$ color-color plane. 
As shown in previous studies on object distributions in optical and mid-infrared 
color-color planes \citep*[][]{eisenhardt2004,rowanrobinson2005}, 
stars and galaxies are more distinctly separated in optical-mid-infrared 
color-color planes than in purely optical color-color planes 
because of the very different temperatures of stars and dust. 
Figure \ref{fig:colorcolor} shows color-color diagrams of the variable objects 
for $R-i'$ versus $i'-z'$ in the bottom left panel and 
$R-i'$ versus $R-m_{3.6\mu\rm{m}}$ in the bottom right panel. 
A galaxy sequence extends widely from bottom to top in the blue side of $R-i'$ colors 
in the $R-i'$ versus $R-m_{3.6\mu\rm{m}}$ color-color plane 
while stellar sequence shows sharp distribution below the line in the figure. 
We defined variable objects satisfying a color criterion, 
$R-m_{3.6\mu\rm m}<2.6\times(R-i')-0.5$, as stars. 

\subsubsection{Results of Star Selection}\label{sec:criteriastar3}
Finally, we combined two independent criteria for stars 
and classified the variable objects into four categories; 
\begin{itemize}
  \item[1)] {\it reliable stars}: variable objects which are selected as stars a) 
and have stellar optical-mid-infrared colors ($78$), 
  \item[2)] {\it probable stars}: variable objects which are selected as probable stars 
b-1) or b-2) and have stellar optical-mid-infrared colors ($24$), 
  \item[3)] {\it possible stars}: variable objects which are selected as possible stars c) 
and have stellar optical-mid-infrared colors ($5$), 
  \item[4)] {\it non-stellar objects}: variable objects which do not have stellar 
optical-mid-infrared colors or are categorized in non-stellar objects d) ($921$). 
\end{itemize}
All the objects in 1), 2), and 3) satisfy the optical color-magnitude critetion ($B-V>0.08\times V-1.59$). 
The numbers of selected variable objects in each selection are given between parentheses. 
In Figure \ref{fig:colorcolor}, filled circles show 1) {\it reliable stars} and 
open circles show 2) {\it probable stars} and 3) {\it possible stars}. 

We have three spectroscopically identified variable stars; an early-type star and two M dwarf stars. 
All these stars are securely classified as stars. 
The early-type star and one M dwarf star are classified as 1) {\it reliable stars}. 
Another M dwarf star is classified as a 2) {\it probable star} with a red color of $R-i'=1.14$, 
a high {\it morphological score} of $m_{\rm score}=9$ (i.e. a point source), and 
a low color score of $c_{\rm{score}}=0.34$. 

The stellar and galaxy sequences in the $R-i'$ versus $R-m_{3.6\mu\rm{m}}$ color-color 
diagram join together around $R-i'\sim0.2$ and $R-m_{3.6\mu\rm{m}}\sim0$. 
Objects in this region can be contaminated with galaxies. 
There are several spectroscopically identified galaxies in the region, 
although they are not recognized as stars due to their optical extended morphologies. 
Point sources showing optical variability can be luminous quasars, 
but optical-mid-infrared colors of quasars are red enough to be discriminated from stars. 
All the 35 SDSS quasars at $0.2<z<3.7$ \citep{hatziminaoglou2005} 
and 259 SDSS quasars at $0.1<z<5.2$ \citep{richards2006} 
have redder optical-mid-infrared colors 
($r-m_{3.6\mu\rm{m}}>0.35$, shown as a large ellipse) 
as well as quasars in our sample than stars with similar optical colors. 
We also examined the robustness of star selections by comparing the star 
count in $V$-band with that predicted by the Besan\c{c}on model. 
Stars in the SXDF were selected from the catalogs in the same criteria as for the variable objects. 
Star counts including {\it reliable stars}, {\it probable stars}, and {\it possible stars} 
in the SXDF reasonably agree with the model prediction in $21<V<24$. 
In the bottom right panel of Figure \ref{fig:colorcolor}. there can be seen several 
$R-i>0.6$ point sources with stellar optical-mid-infrared colors and 
low color scores $c_{\rm{score}}$ on the stellar sequence (gray circles below the dot-dashed line). 
These low color scores $c_{\rm{score}}$ may derive from non-simultaneous observations
\footnote{For example, V-band observations have been mainly carried out after those in other broad bands 
were almost finished \citep{furusawa2007}.} and their variability. 
In this way, our classifications based on colors of objects could be inaccurate 
because of non-simultaneous observations, 
but many facts described above indicate that variable stars are securely selected 
through the criteria with small contaminations and high completeness. 
The {\it reliable stars}, {\it probable stars} and {\it possible stars}, $107$ objects 
in total ($0.889$ deg$^2$, $120$ objects deg$^{-2}$), 
make up our sample of variable stars in subsequent discussions. 

\subsection{SN/AGN Separation}\label{sec:snagnseparate}
After we selected variable stars, 
we classified non-stellar variable objects into SNe and AGN based on two parameters from the optical imaging data: 
the offset of the location of variable component from the host object and the light curve. 

\subsubsection{Criteria on Variable Location}\label{sec:criteriasnagn1}
First, we examined the offsets between the variable components and their host objects. 
Variability should be observed at the centers of the host objects for AGN as well as stars, 
while SNe can explode at any position relative to the host objects. 
In order to estimate the errors of the locations of variable components for AGN, 
we examined the offsets for optically variable X-ray sources. 
Most of the X-ray sources are considered to be AGN. 
Some of them can be stars emitting X-ray, but for our purposes, that is not a problem: 
variable stars also should show variability at their centers. 
Almost all of the X-ray detected variable objects ($\sim97\%$) have spatial offsets 
below $1$ pixel ($0\farcs202$), and the offsets distribute with a scatter 
of $\sigma_{\rm{offset,X-ray}}=0.29$ pixel. 
We also examined offsets for simulated variable objects 
used in calculations of detection completeness in \S\ref{sec:completeness}, 
and found that their offsets from the located positions range with a scatter of 
$\sigma_{\rm{offset,sim}}=0.51$ 
pixel, slightly larger than that of X-ray sources. 
We set the threshold between 
{\it central variability} (variability at the central positions)
and {\it offset variability} (variability at the offset positions) 
to $1.2$ pixel, 
which is $3\sigma_{\rm{offset}}$ (using the average of these two error estimates). 

\subsubsection{Criteria on Light Curve}\label{sec:criteriasnagn2}
We also used light curves to discriminate SNe from AGN. 
Bright phases where we detect object variability should be limited to 
within one year for SNe for two reasons. 
First, multiple SNe very rarely appear in a single galaxy within our observational baselines. 
Second, any SNe should become fainter than our detection limit within one year after 
the explosions in the observed frame. 
SNe Ia, the brightest type except for hypernovae and some of type-IIn SNe, can be seen 
at $z\sim1.4$ at our detection limit ($i'_{\rm{vari}}\sim25.5$ mag). 
One year in the observed frame corresponds to five months in the rest frame 
at $z=1.4$ due to the time dilation by cosmological redshift. 
Therefore, SNe should fade by more than $4$ mag below the maximum brightness \citep[e.g.,][]{jha2006}. 
This decrease of brightness is a minimum value of observed declining magnitudes of SNe 
and is almost the same as the dynamic range of our detection for variability ($i'_{\rm{vari}}=20.5-25.5$ mag). 
Hence, we defined objects which are at bright phases for less than one year, 
and are stably faint in the remainder of the observations, as 
''objects with SN-like light curves''. 
In order to apply this criterion for the variable objects, 
it is necessary for light curves to cover more than three years at least. 
In two fields (SXDF-N and SXDF-W), our images cover only two years, 2002 and 2003; 
therefore, variable objects in these two fields can not be classified as like or unlike SNe. 
Hence, we concentrate on the other three fields (SXDF-C, SXDF-S, and SXDF-E) 
to discuss number densities of SNe and AGN based on robust classifications. 
Among $693$ variable objects in these three fields, $619$ objects are classified as non-stellar objects. 
Out of $619$ non-stellar variable objects, light curves of $27$ objects 
cover only two years because of insecure subtractions in some epochs. 
Therefore, we use $592$ objects in the following section when 
separating SNe from AGN. 
The number densities derived below can increase 
by a factor of $1.05 (=619/592)$ to account for the objects lost to poor subtractions. 
We also note that these criteria provide mis-classifications for objects with light curves of 
poor signal-to-noise ratios. 

\subsubsection{Results of SN/AGN Separation}\label{sec:criteriasnagn3}
We combined these two criteria; the position offsets and the light curves, 
and applied them for $592$ objects over $0.566$ deg$^2$ to classify variable objects as SNe and AGN. 
Variable objects with significant ($>1.2$ pixel) offsets and SN-like light curves 
(case 1: $158$ objects) are highly likely to be SNe, while those with small 
($\leq1.2$ pixel) offsets and non-SN-like light curves 
(case 2: $228$ objects) are highly likely to be AGN. 
There are no clear contaminations in each case in our sample; 
we have no X-ray sources or spectroscopically identified broad-line AGN classified as case 1, 
and also have no SNe with spectroscopic identifications or 
spectroscopic redshift determinations of their host galaxies in \citet{yasuda2007} classified as case 2. 

Objects with small offsets and SN-like light curves 
(case 3: $160$ objects) or significant offsets and non-SN-like light curves 
(case 4: $46$ objects) are hard to judge. 
In our time samplings and baselines, AGN can be classified as SNe in terms of light curves 
and objects in case 3 can be either SNe or AGN. 
Out of $14$ SNe with spectroscopic identifications or spectroscopic redshift determinations 
of their host galaxies, 
$8$ objects are classified as case 1 (significant offsets and SN-like light curves) 
and $5$ objects are classified as case 3 (small offsets and SN-like light curves). 
Only 1 object are classified as case 4 (significant offsets and non-SN-like light curves) 
probably because the variable component of this object is the faintest 
among these $14$ SNe ($i'_{\rm{vari}}=25.3$mag). 
Through our criterion for offsets of variability, $36\%\ (=5/14)$ of SNe are recognized 
as {\it central variability}, although spectroscopic observations for SNe should be biased 
to {\it offset variability} to avoid AGN. 
If we use this ratio, about $88\ (=158\times5/9)$ of $160$ objects in case 3 
are considered to be SNe. 
On the other hand, among $644$ variable objects in these fields whose light curves 
covering longer than three years, the number of X-ray sources is $128$. 
Out of $128$ optically variable X-ray sources -- which we believe are AGN -- 
$89$ objects are in case 2, $37$ objects are in case 3, $1$ object is in case 1, 
and $1$ object is in case 4. 
Thus, $30\%\ (=(37+1)/128)$ of AGN have SN-like light curves. 
If we assume that this same fraction of all AGN have SN-like light curves, 
then about $96\ (=228\times(37+1)/(89+1))$ of $160$ objects in case 3 are considered to be AGN. 
The expected total number of these SNe and AGN in case 3 is $184\ (=88+96)$ and 
reasonably consistent with the real number ($160$) in case 3. 
Hence, we made number densities of SNe and AGN by scaling those for well-classified samples 
(case 1 for SNe and case 2 for AGN) by factors of $14/8$ for SNe and $128/89$ for AGN, 
as shown in Figure \ref{fig:varinum_time1} for the host object magnitudes and 
Figure \ref{fig:nc_varicomp} for the variable component magnitudes. 
We discuss these figures in \S\ref{sec:overall_nc}. 
We obtained these number densities of $489$ and $579$ objects deg$^{-2}$ 
for SNe and AGN, respectively. 

Nature of objects in case 4 is mysterious. 
We have only two spectroscopic data for objects in case 4 and both of the two spectra 
indicate that they are normal galaxies at $z\sim0.5$. 
Their offsets from the centers of the host objects are significant. 
The light curves have marginal parameters around the threshold between those of SN-like light curves and 
non-SN-like light curves, and they may be misclassified as non-SN-like light curves due to measurement errors. 
On the other hand, only one object in case 4 is detected in X-ray. 
This object and some others have offsets from the centers of their host galaxies 
just above the threshold ($1.2$ pixel). 
Another possible reason why some objects are classified as case 4 is 
misidentifications of the host objects due to their faintness. 
Such objects might be also transient objects. 
The number of objects in case 4 is not large and we do not include them for discussions of number densities. 

\section{Detection Completeness of Optically Variable Objects}\label{sec:completeness}
We intensively studied two types of completeness in this paper. 
One is completeness of variability detection itself, which should be 
functions of PSF sizes and background noise of the images, and 
should be independent of properties of object variability. 
Another completeness depends on behaviors of object variability and observational time samplings. 
It is not easy to estimate this type of completeness because properties of variability 
are complicated and differ from object to object. 

\subsection{Variability Detection Completeness}\label{sec:detectionitself}
\begin{figure}[htbp]
\begin{center}
\epsscale{.8}
\includegraphics[angle=270,scale=0.3640]{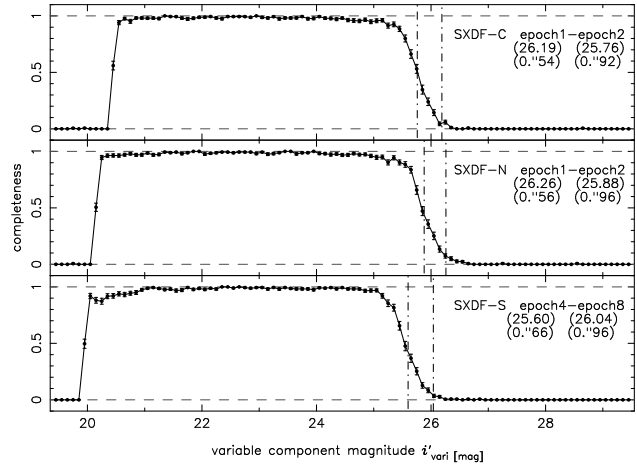}
\caption{Detection completeness of variable objects as a function of variable component 
magnitude $i'_{\rm{vari}}$ calculated by adding artificial objects in the images 
and detecting them in the same manner as for the real images. 
Three cases are shown in the figure. 
The limiting magnitudes of the images before subtractions are given between the top parentheses 
and drawn in dot-dashed lines. 
FWHM of PSF are also described between the bottom parentheses. 
Cut-offs at bright magnitudes are caused by masking bright objects 
in order that our detection are not affected by saturation or non-linearity of the CCD pixel. 
\label{fig:completeness}}
\end{center}
\end{figure}

The first type of completeness is simply estimated 
by locating artificial variable objects (point sources with PSFs of the same size as those in the real images) 
randomly in the stacked images using the {\it artdata} task in IRAF, 
and detecting them in the same manner as for the real images. 
Three examples 
are shown in Figure \ref{fig:completeness}. 
We can almost completely detect variable objects whose flux differences in magnitude unit 
are brighter than $\sim25.0-25.4$ mag. 
The cut-offs at the bright ends around $i'_{\rm{vari}}=20.0-20.5$ mag are caused by masking bright objects 
to avoid false detections of variability. 
The detection completeness decreases down to zero at $i'_{\rm{vari}}\sim26.0-26.3$ mag. 
The shapes of completeness curves are similar for all the subtracted images, 
although there are slight offsets in magnitude axis due to differences of 
limiting magnitudes of the stacked images. 
The completeness is almost uniform in the whole regions of the subtracted images 
where we investigate object variability. 

\subsection{Detection Completeness for Variable Stars, SNe, and AGN}\label{sec:completenessstaragn}
Variability detection completeness depends on shapes of light curves of variable objects 
and observational time samplings. 
We want to know detection completeness for each class of variable objects. 
It is difficult to do so for objects showing burst-like variability 
such as dwarf stars, SNe, and AGN, while we can easily simulate light 
curves of pulsating variable stars showing periodic variability. 
AGN optical variability has been often characterized by the structure function 
\citep*{kawaguchi1998,hawkins2002,devries2003,vandenberk2004,devries2005,sesar2006}, 
and we can simulate AGN light curves to estimate the detection completeness. 
Hence, we consider two types of simulated light curves; periodic variability for 
pulsating variable stars and variability characterized by the structure functions 
for AGN and estimate completeness. 
We also evaluate variability detection efficiency using real light curves and the X-ray source sample for AGN. 

In our time samplings and depths, SNe Ia up to $z\sim1.4$ can be detected in efficiency of a few tens of percent. 
Detection efficiency of SNe will be examined in detail in our following paper on SN rate. 

\subsubsection{Periodic Variability}\label{sec:period}
\begin{figure}[htbp]
\begin{center}
\epsscale{.8}
\includegraphics[angle=270,scale=0.30]{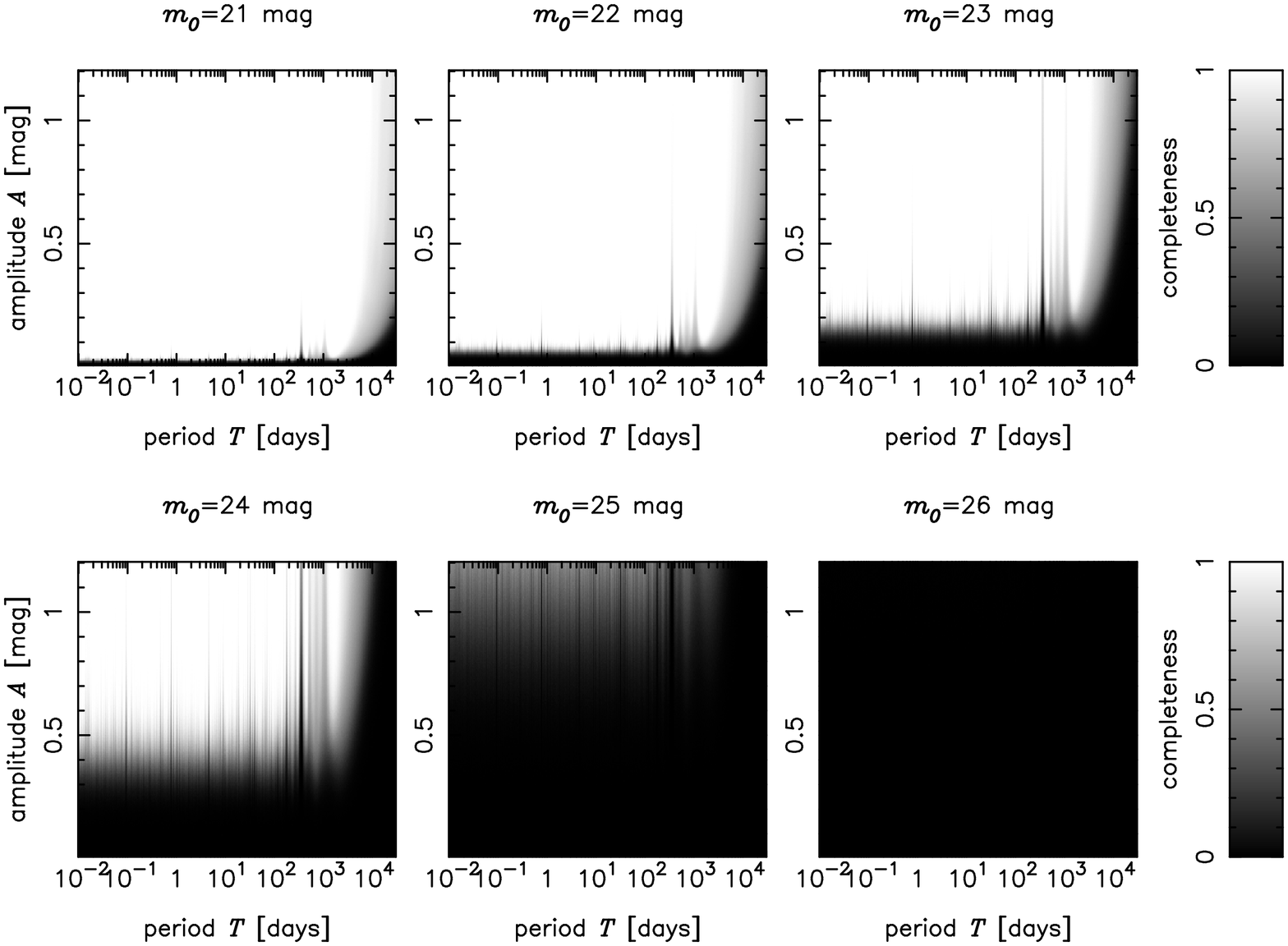}
\includegraphics[angle=270,scale=0.30]{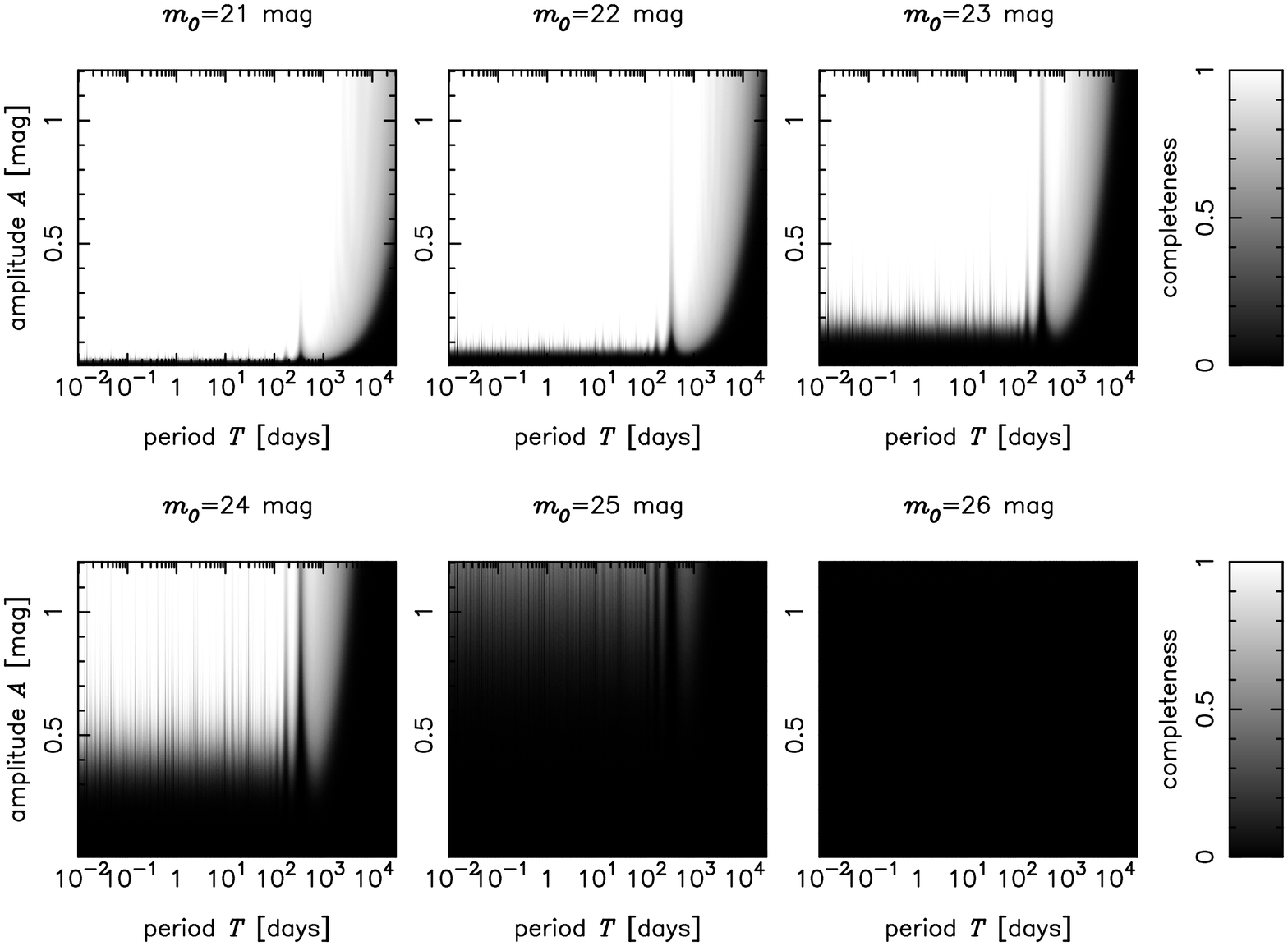}
\caption{Detection completeness for variable objects showing periodic variability with periods $T$, 
amplitude $A$, and averaged magnitude $m_0$ in SXDS-C (10 epochs from 2002 to 2005, upper 6 panels) 
and SXDS-N (8 epochs from 2002 to 2003, lower 6 panels). 
\label{fig:complvari_1}}
\end{center}
\end{figure}

Detection completeness for pulsating variable stars whose variability are 
periodic is easily calculated. 
Simulated light curves are parameterized by magnitude amplitudes $A=0.0-1.0$ mag, 
periods $T=0.01-20000$ days, and averaged magnitudes $m_0=21-26$ mag
\footnote{Light curves are expressed as $m(t)=\frac{A}{2}\sin(2\pi t/T)+m_0$.}. 
We have roughly two types of time baselines of observations, 
and then we show the completeness for the two fields, 
SXDF-C with observations from 2002 to 2005, and SXDF-N with those from 2002 to 2003, 
in Figure \ref{fig:complvari_1}. 
There are slight differences between the detection completeness for the two fields, 
but, both the completeness are almost the same on the whole. 
In all the five fields, some dark spikes indicating low sensitivity to those periods 
can be seen because our observations have been carried out only in fall. 
The completeness is a strong function of magnitudes and amplitudes. 
We can detect only variable stars with large amplitudes ($A>0.4$ mag) for faint stars of $i'>24$ mag. 

\subsubsection{AGN Variability}\label{sec:variagn}
Unlike pulsating stars, AGN usually show burst-like variability aperiodically. 
Behaviors of AGN variability depends on rest-frame wavelength. 
Generally, AGN variability are larger in shorter wavelength. 
Considering the time dilation of cosmological redshift, 
detection efficiency of AGN by optical variability in a certain 
broad band, in $i'$-band in this study, is not clear. 
To the zeroth approximation, effects of wavelength dependence and time dilation 
are cancelled each other. 

Quasar variability has been often described in the form of the structure function, 
\begin{eqnarray}
SF(\Delta{t}){\equiv}\left[{\sum}\left\{m(t+{\Delta}t)-m(t)\right\}^2
/N(\Delta{t})\right]^{0.5}, 
\end{eqnarray}
where $N(\Delta{t})$ is the number of objects with time intervals of $\Delta{t}$. 
As far as we focus on time scales of months to several years (not several decades), 
a power-law form of the structure functions, $SF(\Delta{t})=(\Delta{t}/\tau_0)^\gamma$, 
parameterized by a characteristic time scale of variability, $\tau_0$, 
and a power-law slope, $\gamma$, is well fitted to observational data. 
The structure functions begin to show a turnover around the rest-frame time lag of a few years 
and this power-law approximation slightly overestimates the variability \citep*{ivezic2004b,devries2005,sesar2006}.
However, our observational data span over only three years at longest in observed frame 
and this simplification is good enough for our estimation of AGN detection completeness. 
\begin{figure}[htbp]
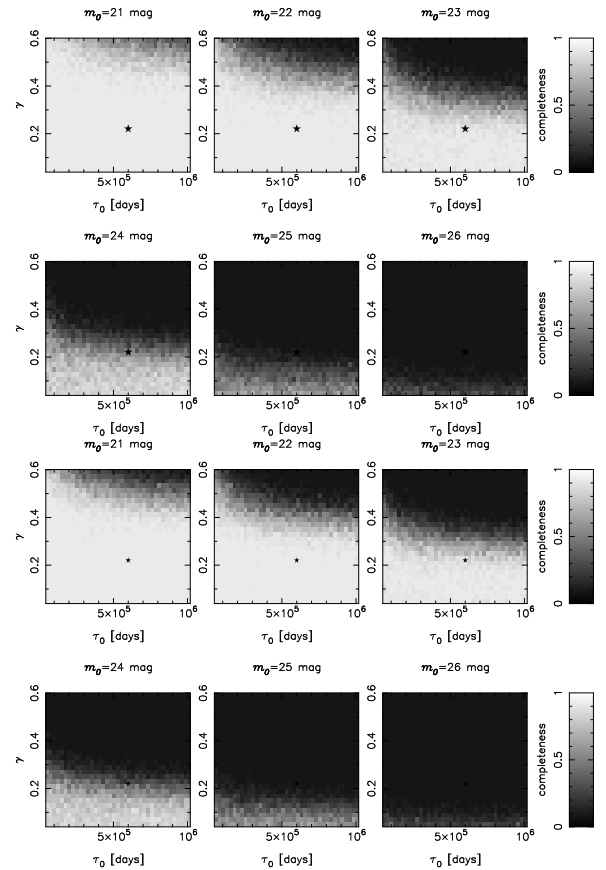

\begin{center}
\epsscale{.8}
\includegraphics[angle=270,scale=0.30]{f10a.eps}
\includegraphics[angle=270,scale=0.30]{f10b.eps}
\caption{Detection completeness for variable objects whose variability are characterized 
by the structure functions 
in SXDS-C (10 epochs from 2002 to 2005, upper 6 panels) 
and SXDS-N (8 epochs from 2002 to 2003, lower 6 panels). 
Typical values of $\gamma$ and $\tau_0$ for quasars are $0.2$ and $6\times10^5$ days, respectively, 
these values are plotted as stars. 
\label{fig:complvari_3}}
\end{center}
\end{figure}

The slope of structure function provides us important keys to the origin of their variability 
\citep{hughes1992,hawkins1996,kawaguchi1998,hawkins2002}; 
accretion disk instability \citep{rees1984}, 
bursts of SN explosions \citep{terlevich1992}, 
and microlensing \citep{hawkins2002}. 
Therefore, the slope of structure function has been investigated in many previous studies. 
Monitoring observations of Palomar-Green (PG) quasars \citep*[][]{giveon1999,hawkins2002} and variability 
studies of enormous number of SDSS quasars between SDSS imaging data and older plate imaging data 
\citep*{devries2003,devries2005,sesar2006}, or SDSS spectrophotometric data 
\citep{vandenberk2004} set constraints on characteristic behaviors of variability. 
They obtained slopes of $\gamma\sim0.2$. 
We simulated light curves satisfying the structure functions 
with $5\times10^4<\tau_0<1\times10^6$ days and $0.05<\gamma<0.6$ 
for averaged apparent magnitudes $m_0=21-26$ mag for AGN components, 
and calculated detection completeness in our observational time samplings. 
We show results in Figure \ref{fig:complvari_3} for two cases 
(4-year baseline for SXDF-C, -S, and -E and 2-year baseline for SXDF-N and -W) 
done for periodic variability in \S\ref{sec:period}. 
Values of the slope obtained in previous studies indicate that 
main origin of AGN optical variability is accretion disk instability or microlensing 
and the typical values of $\tau_0$ and $\gamma$ 
are $\sim6\times10^5$ days and $\sim0.2$ in observed frame, respectively, 
which are indicated as stars in Figure \ref{fig:complvari_3}. 
These values were derived from observations in bands bluer than $i'$-band which is used in this work. 
In $i'$-band, it is expected that $\gamma$ stays roughly constant and $\tau_0$ become slightly larger. 
Determination errors of $\tau_0$ and $\gamma$ are not small here 
and we infer that the completeness for our data is not less than $\sim50\%$ for $m_0<24$ mag. 

\begin{figure}[htbp]
\begin{center}
\epsscale{.8}
\includegraphics[angle=270,scale=0.344]{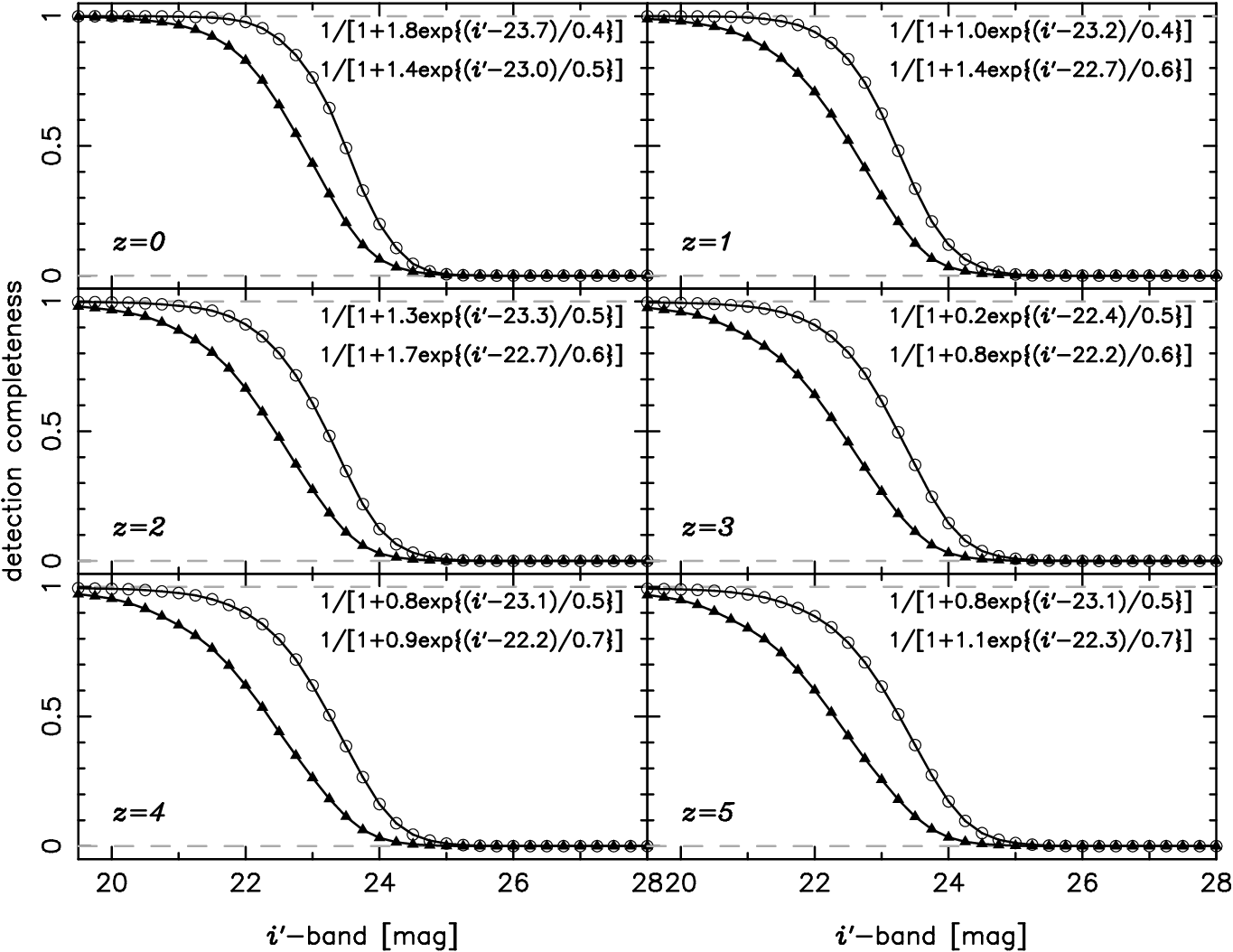}
\caption{Detection completeness for $42$ PG quasars using real light curves over seven years 
obtained by \citet{giveon1999}. 
In the cases of objects at $z=0$, $1$, $2$, $3$, $4$, and $5$, 
we calculated the detection completeness considering cosmological time dilation 
and variability dependence on rest-frame wavelength \citep{vandenberk2004}. 
We show the results for time samplings in SXDF-C in open circles and 
for those in SXDF-N in filled triangles. 
The fitted lines are $1/[1+a\times\exp\{({\rm{mag}}-b)/c\}$, parameterized by $a$, $b$, and $c$. 
\label{fig:complvari_giveon1999varidep}}
\end{center}
\end{figure}

The detection completeness for quasars were also estimated using observational light curves 
of $42$ PG quasars in $B$ and $R$-bands over seven years obtained by \citet{giveon1999}. 
The photometric points were well sampled and their observational time baselines are much longer than ours. 
They lacked time samplings in time scales of days and we interpolated the light curves. 
Their quasars are at relatively low redshift ($0.061<z<0.371$) with luminosity of $-26.3<M_B<-21.7$ mag. 
The well-sampled light curves are very useful for calculating the completeness in our survey. 
In our estimations of the detection completeness, we assume the photometric errors of our surveys 
considering the time dilation and dependence of variability on rest-frame wavelength which was indicated 
in \citet{vandenberk2004} using the SDSS quasars. 
Variability depends on rest-frame wavelength that quasars are 
about twice as variable at $1000$\AA\ as at $6000$\AA\ as shown in their Equation 11 
\footnote{variability $v(\lambda) \rm{[mag]}=0.616\exp(-\lambda/988\rm{\AA})+0.164$.} and Figure 13. 
We obtained detection completeness curves as a function of observed $i'$-band magnitude as shown in Figure \ref{fig:complvari_giveon1999varidep}. 
From top left to bottom right, the redshifts are $z=0$, $1$, $2$, $3$, $4$, and $5$. 
The results for the two fields, SXDF-C from 2002 to 2005 and SXDF-N from 2002 to 2003, are plotted. 
There are some differences between these cases and the longer baselines give us higher completeness. 
In this simulation, our detection completeness for AGN is down to zero at $\sim24.5-25.0$ mag. 
Redshift dependence of the completeness are very small due to cancellation of time dilation and 
dependence of variability on rest-frame wavelength. 
We note that dependence of variability on AGN luminosity are not considered. 
In our detection limit, we can observe Seyfert-class AGN at low redshift and 
our estimates of completeness here may be lower limits for them 
because of the anti-correlation between AGN luminosity and variability. 

Optical variability of $172$ X-ray sources are detected, while 
we have $481$ X-ray sources in our variability survey field. 
The fraction of objects showing optical variability among X-ray sources is $36\pm2\%$. 
All of the X-ray detected variable objects are brighter than $i'=24.4$ mag. 
Then, we also limit X-ray sources to those with $i'<24.4$ mag, the number is $334$. 
Simply assuming that all the X-ray sources can show large variability enough to 
be detected, the detection completeness for X-ray sources is $51\pm3\%$. 
\citet{vandenberk2004} found that X-ray detected quasars show larger variability 
than quasars which are not detected in X-ray. 
This tendency is true for our sample; 
X-ray brighter sources show larger optical variability 
than X-ray fainter sources in a certain optical magnitude. 
Variable objects not detected 
in X-ray should have lower completeness than variable objects detected in X-ray. 
Some of the X-ray sources are type-2 populations whose optical variability are 
more difficult 
to be detected given the unified scheme of AGN and 
detection completenss of type-1 AGN can be higher than these estimates. 

These three estimations of detection completeness for AGN are roughly consistent with each other 
and strongly depend on apparent magnitudes as well as that for pulsating variability. 
Inferred completeness is $\sim1$ at $i'\sim21$ mag and slowly decreases down to zero at $i'=24-25$ mag
with uncertainty of a factor of a few. 

\section{Overall Sample}\label{sec:overall_nc}
\begin{figure}[htbp]
\begin{center}
\epsscale{.8}
\includegraphics[angle=270,scale=0.384]{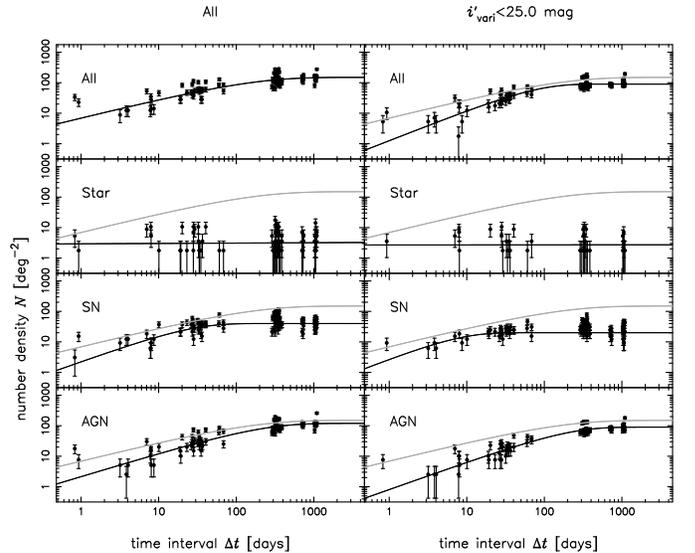}
\caption{Number densities of variable objects as a function of 
time intervals of $\Delta{t}$ in $i'$-band imaging observations. 
We plot them for objects showing variability above $5\sigma_{\rm{lc}}$ in the left column, 
objects with variable component magnitude $i'_{\rm{vari}}$ brighter than $25.0$ mag in the right column. 
These points are fitted in the form of $N=a\times\exp\{-b(\Delta{t})^c\}$ (solid lines). 
The fitted line in the top left panel is shown in gray lines in every panel as a reference. 
\label{fig:varinum_time1}}
\end{center}
\end{figure}

We found $1040$ variable objects including possible transient objects 
in $i'$-band in the SXDF down to 
variable component magnitudes of $i'_{\rm{vari}}\sim25.5$ mag. 
In the top left panel of Figure \ref{fig:varinum_time1}, the number densities $N$ 
of all the detected variable objects are shown 
as a function of time interval $\Delta{t}$ of observations 
when we compare only two separated images. 
We also draw a line fitted in the form of 
$N(\Delta{t})=a[1-\exp\{-b(\Delta{t})^c\}]$ 
Number densities as a function of $\Delta{t}$ for SNe are expected to bend over 
at a certain time interval because variability time scales of SNe are roughly common to each other, 
a few months, in rest-frame. 
Then, we introduce such function forms for all the kinds of variable objects. 
Detection limits for variable objects are slightly different for each combination 
of observational epochs and it can affect obtained number densities. 
Then, we also show number densities of variable objects with variable components 
brighter than $25.0$ mag in the right panels. 
Monotonic increases of number densities in observed frame 
indicate that objects showing variability in time scales of years are 
dominant in this magnitude range (and in the SXDF direction; the Galactic halo). 
The number densities for variable stars are shown 
in the second row, SNe in the third row, and AGN in the bottom row 
after classifying variable objects in $\S$\ref{sec:objclass}.
Variable objects for calculating number densities used in this figure are $644$ objects with light curves 
in long baselines from 2002 to 2005 enough to discriminate SNe from AGN, and are within the IRAC field. 
These SN and AGN number densities are corrected for case 3 objects. 
Typical time scales of variability are from days to months for variable stars and SNe 
while AGN show variability in time scales of months to years \citep[e.g.,][]{vandenberk2004}. 
The flat distributions for variable stars indicate that 
stars showing variability in short time scales are mainly included in the sample. 
The SN number densities arrive at plateau in time scales of about a few months as expected from 
time scales of SN light curves. 
On the other hand, significant increases towards time scales of years are seen for AGN. 
Fitted lines for AGN in the same form of function as for SNe show possible turnovers 
at $\Delta{t}\sim1000$ days. 
Results of SDSS quasars \citep{ivezic2003} indicated turnovers of structure functions 
at $\Delta{t}\sim1000$ days in rest-frame and these possible turnovers in our results 
may be real. 
These results indicate that long time baselines of years make the completeness 
significantly higher only for AGN. 

\begin{figure}[htbp]
\begin{center}
\epsscale{.8}
\includegraphics[angle=270,scale=0.33124]{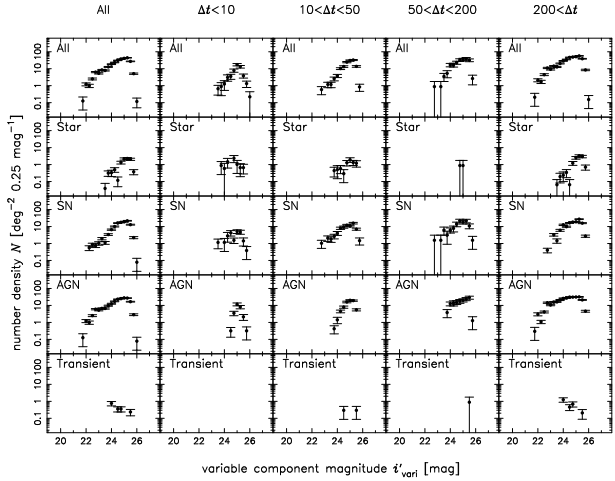}
\caption{Number densities of variable objects as a function of variable component magnitude $i'_{\rm{vari}}$. 
We plot those for all the variable objects in the top row, 
variable stars in the second row, 
SNe in the third row, and AGN in the bottom row, respectively. 
Number densities in each time scale, $\Delta{t}<10$ days in the second column, 
$10<\Delta{t}<50$ days in the third column, 
$50<\Delta{t}<200$ days in the fourth column, 
and $\Delta{t}>200$ days in the right column, 
as well as in the case using all the observational data in the left column, are also plotted. 
Number densities of possible transient objects are shown in the bottom row. 
\label{fig:nc_varicomp}}
\end{center}
\end{figure}

We show the number densities of variable objects as a function of 
variable component magnitude $i'_{\rm{vari}}$ 
in the left column of Figure \ref{fig:nc_varicomp}. 
Number densities after classifying variable objects into variable stars, 
SNe, and AGN in three time scales, $\Delta{t}<10$ days in the second column, 
$10<\Delta{t}<50$ days in the third column, 
$50<\Delta{t}<200$ days in the fourth column, 
and $\Delta{t}>200$ days in the right column, are also plotted. 
All the number densities drop around $i'_{\rm{vari}}\sim25$ mag and 
our variability detections are reasonably consistent with simulated 
completeness as shown in Figure \ref{fig:completeness}. 
In the bottom row of Figure \ref{fig:nc_varicomp}, the number densities of 
transient objects are plotted. 
By definition of transient objects (see \S\ref{sec:objclass}) , 
variable component magnitudes of them 
correspond to total magnitudes in their brighter phases. 
The transient object sample can include not only objects showing 
transient phenomena such as flare-ups of 
faint dwarf stars but also less slowly moving Kuiper belt objects 
than $\omega\sim1''$ hour$^{-1}$ (a typical value of seeing size per 
exposure time for each stacked image), corresponding to a semimajor 
axis of $>100$ AU. 
These estimates of numbers of variable objects provide us the expected numbers of variable objects 
contaminated into interested samples using non-simultaneous observational data. 
For example, \citet{iye2006} and \citet{ota2007} investigated 
the possibility of variable object contaminations 
into their LAE sample at $z\sim7.0$ in the Subaru Deep Field \citep[SDF;][]{kashikawa2004}, 
which were obtained comparing broad-band imaging data taken before 2004 
with narrow-band NB973 imaging data in 2005. 
From the bottom row of Figure \ref{fig:nc_varicomp}, a few or less 
than one objects in the SDF ($\sim0.3$ deg$^2$) 
can be just transient objects and misclassified as narrow-band excess objects in their sample. 
Their plausible candidates with narrow-band excess are two. 
One of them were spectroscopically identified and turned out to be a real LAE at $z=6.94$. 
Spectroscopic identification for another candidate should be done. 
Whether this candidate is a real LAE or not, the number of narrow-band excess objects they found is 
consistent with expected number from the statistics of number densities of transient objects. 

\begin{figure}[htbp]
\begin{center}
\epsscale{.8}
\includegraphics[angle=270,scale=0.3741]{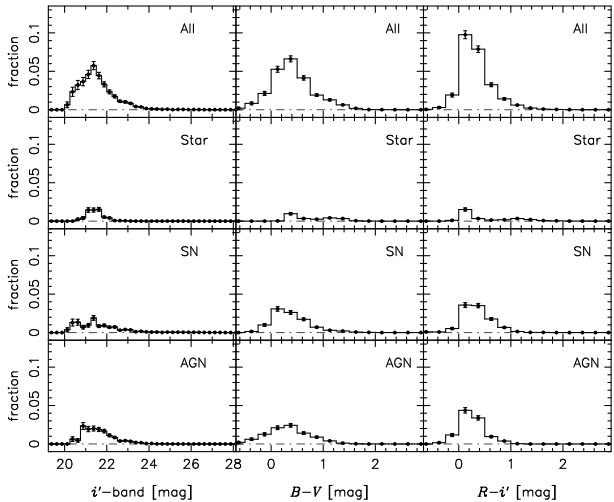}
\caption{Fractions of all the variable objects to the whole objects in the SXDF 
as functions of $i'$-band magnitude, $B-V$, and $R-i'$ colors in the top row. 
Fractions after classifying variable objects are also plotted in the second, third, and bottom rows. 
\label{fig:fracvari}}
\end{center}
\end{figure}

The fractions of variable objects to the overall objects in the SXDF are shown 
in Figure \ref{fig:fracvari}. 
Magnitudes used in this figure are the total magnitudes of host objects. 
About $5\%$ of objects at $i'\sim21$ mag show optical variability, 
and the fractions rapidly decrease towards fainter magnitudes. 
These declines are caused by detection limit for variability. 
Large variability relative to the host objects are necessary to be detected for faint objects. 
Almost all of objects in the SXDF are galaxies, not stars (the fraction of stars is $\sim4\%$). 
Then, the fraction of variable AGN is $\sim3\%$ around $i'\sim21$ mag, 
where the detection completeness is $\sim1$. 
\citet{sarajedini2006} found that $2.6\%$ of galaxies have variable nuclei 
down to $V_{\rm{nuc}}<27.0$ mag without completeness corrections 
using two-epoch observations separated by seven years. 
Their fraction is consistent with ours. 
\citet{cohen2006} also used four-epoch ACS imaging data in time baselines of three months 
down to $V_{\rm{total}}<28$ mag and detected variability of $\sim1\%$ of galaxies. 
This small percentage in \citet{cohen2006} is consistent with ours 
if we consider their short time baselines because most of AGN vary in brightness in longer time scales 
of years as shown in Figure \ref{fig:varinum_time1}. 

\section{Variable Stars}\label{sec:varistar}
\begin{figure}[htbp]
\begin{center}
\epsscale{.8}
\includegraphics[angle=270,scale=0.34]{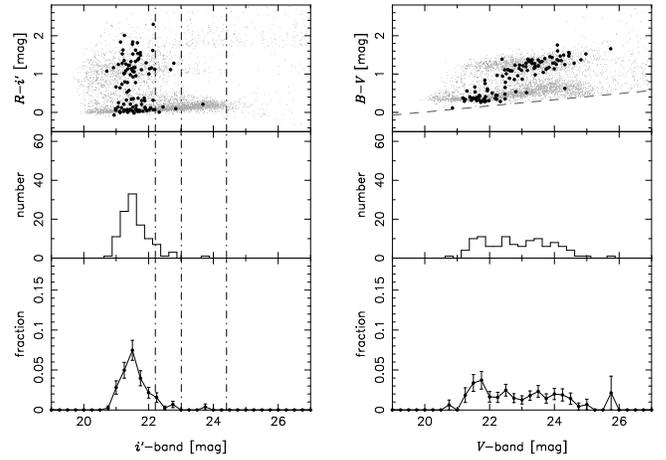}
\caption{Top panels show distributions of variable objects (black circles) in color-magnitude 
diagrams, $V$ versus $B-V$ in the left panel and $i'$ versus $R-i'$ in the right panel. 
Non-variable stars are also shown in gray dots. 
The criterion for excluding contaminations of galaxies, $B-V>0.08\times V-1.59$, 
is drawn in a dotted line in the $V$ versus $B-V$ diagram. 
Number counts and fractions of variable objects for $i'$ and $V$ magnitudes are shown 
in the second and third rows, respectively. 
\label{fig:colormaghist}}
\end{center}
\end{figure}

The sample of variable stars used in this section includes $153$ objects 
which were classified as 1) {\it reliable stars}, 2) {\it probable stars}, 
and 3) {\it possible stars} in \S\ref{sec:starselect}. 
The top panels of Figure \ref{fig:colormaghist} show color-magnitude diagrams 
of $i'$ versus $R-i'$ in the left panel and $V$ versus $B-V$ in the right panel 
for variable stars (black circles) and non-variable stars in the SXDF (gray dots). 
\begin{figure}[htbp]
\begin{center}
\epsscale{.8}
\includegraphics[angle=270,scale=0.34]{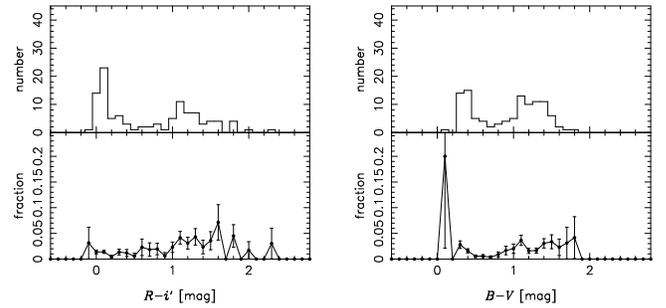}
\caption{Histograms of numbers of variable objects are shown in the upper panels. 
The fractions are also plotted in the lower panels. 
The left column is for $R-i'$ color and the right column is for $B-V$ color. 
\label{fig:colormaghist2}}
\end{center}
\end{figure}
These non-variable stars are selected through the same criteria as for the variable stars. 
Number counts and fractions of variable stars are also plotted in the rows below. 
Figure \ref{fig:colormaghist2} shows color distributions and fractions of variable stars 
for $R-i'$ and $B-V$. 
The number count in $i'$-band has a relatively steep peak at $i'\sim21.5$ mag 
and decreases towards fainter magnitudes. 
This cut-off is partly caused by the variability detection limit. 
The dot-dashed lines indicated in the left columns of Figure \ref{fig:colormaghist} 
are total magnitudes of objects with variable components of the detection limit, $i'_{\rm{vari}}\sim25.5$ mag, 
in the cases of variability amplitudes of $0.05$, $0.1$, and $0.3$ mag from left to right. 
Thus, limiting magnitudes can be as shallow as $\sim22.2$ mag for 
variable objects with low variability amplitudes of $\sim0.05$ mag. 
Poor time samplings of our observations prevent us from determining 
amplitudes of variability in cases of periodic variability. 
Assuming that amplitudes roughly equal to magnitude differences of 
objects between their maxima and minima in our samplings, 
amplitudes of almost all of variable stars are less than $0.1$ mag as shown in Figure \ref{fig:amplstar}. 
These variable stars with low amplitudes can be detected above $\sim22.2$ mag. 
Variable stars with larger amplitudes can be detected for fainter stars. 
The top right panel of Figure \ref{fig:amplstar} clearly 
indicates this tendency of variability selection effects. 

Bimodal distributions are clearly seen in the figures of color-magnitude diagrams (Figure \ref{fig:colormaghist}) 
and color distributions (Figure \ref{fig:colormaghist2}); 
there are two populations of blue bright ($V\sim22$ mag) variable stars 
and red faint ($V\sim23.5$ mag) variable stars. 
From comparison with the Besan\c{c}on model prediction, stars in the upper sequence 
belong to the thick disk populations rather than thin disk 
because stellar sequence of the thin disk is expected to be redder than 
the observed sequence by $B-V$ of $\sim0.3$. 
The red variable stars are considered to be dwarf stars 
showing flares or pulsating giant stars. 
On the other hand, stars in the lower sequence belong to the halo populations. 
The blue variable stars are considered to be pulsating variable stars such as $\delta$ Scuti, 
RR Lyrae, and $\gamma$ Doradus stars, 
within the instability strip in color-magnitude diagrams. 
Eclipsing binaries may be also included. 
\begin{figure}[htbp]
\begin{center}
\epsscale{.8}
\includegraphics[angle=270,scale=0.34]{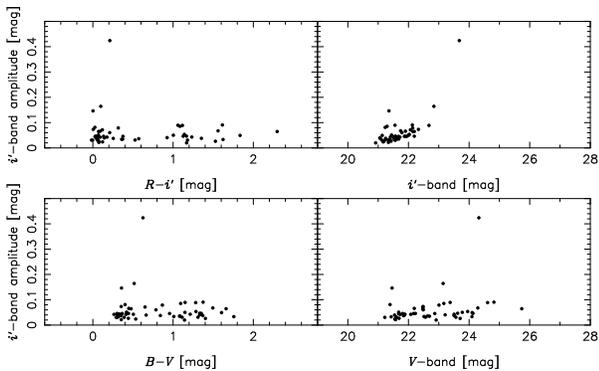}
\caption{Variability amplitudes of variable stars as function of apparent magnitude ($i'$-band and $V$-band) 
and colors ($R-i'$ and $B-V$). 
These amplitudes are defined as magnitude differences between maxima and minima in our time samplings 
and give lower limits of real amplitudes.  
\label{fig:amplstar}}
\end{center}
\end{figure}

\citet{ivezic2000} selected RR Lyrae candidates using SDSS multi-epoch wide-field imaging data 
in short time scales of days, and found a cut-off magnitude at $r^\ast\sim21$ mag. 
They indicated that this cut-off magnitude 
corresponds to a distance of $\sim65$ kpc from the Galactic center, 
which might be the edge of the Galactic halo. 
Our time samplings are too poor to determine their periods of variability as well as real amplitudes. 
Maxima of magnitude differences in our time samplings shown in Figure \ref{fig:amplstar} 
are expected to be smaller than real amplitudes by a factor of $\sim1.2$ from simulated light curves. 
Our typical exposure time of one hour, which is not much longer than variability periods of RR Lyrae 
\citep[$0.3-0.6$ days,][]{vivas2004}, also makes these magnitude differences small. 
Moreover, variability in $i'$-band are smaller than those in $V$-band 
for RR Lyrae \citep[$0.5-1.0$ mag in $V$-band,][]{vivas2004} 
by a factor by $\sim1.5$ \citep[interpolation of values in Table 7 in][]{liu1990}. 
In total, the maxima of magnitude differences in $i'$-band are expected to be less than 
$0.2-0.5$ mag for RR Lyrae in our time samplings. 
Given the variability amplitudes of blue variable stars with $B-V\sim0.3$ in our sample, 
there are one or a few RR Lyrae candidates with amplitudes as large as those of RR Lyrae 
(the bottom left panel of Figure \ref{fig:amplstar}). 
Variability amplitude of the spectroscopically identified blue variable star with $B-V=0.44$ shown 
in the bottom panel of Figure \ref{fig:specvaristar} is small and can not be a candidate of RR Lyrae. 
If these RR Lyrae candidates in our sample are really RR Lyrae, 
the distances from the Galactic center, which are calculated by 
apparent $V$-band magnitude, are larger than $\sim150$ kpc. 
The inferred number densities of RR Lyrae are $\sim10^{-2}$ kpc$^{-3}$ 
at a distance of $\sim150-250$ kpc from the Galactic center towards the halo, 
and consistent with extrapolations of results by \citet{ivezic2000} 
and \citet{vivas2006} in spite of our poor statistics. 
They are candidates of the most distant Galactic stars known so far. 
Other types of blue variable stars in the halo population than RR Lyrae 
are considered to be included in the sample such as Population II $\delta$ Scuti stars, 
$\gamma$ Doradus stars, eclipsing binaries, and so on. 
In order to examine nature of these faint blue variable stars, 
dense monitoring observations for determinations of pulsating periods and amplitudes, 
and/or follow-up spectroscopic observations are necessary. 

Two of the red variable stars with $B-V=1.36$ and $1.34$ were spectroscopically observed 
and identified as M dwarf stars (Figure \ref{fig:specvaristar}). 
The $B-V$ colors of the red faint variable stars are $\sim0.8-2.0$, which are those of K or M type stars. 
The $V$-band absolute magnitudes of these stars are widely spread, for example, $M_V=8-12$ mag 
for M stars with colors of $1.3<B-V<1.7$ in the case of main sequence from the Hipparcos data \citep{koen2002}. 
Assuming that almost all of these red variable stars are dwarf stars, not red giant stars, 
the inferred distances to the stars are $\sim1.5-4$ kpc from the Galactic plane, 
where the thick disk population is dominant \citep{chen2001}. 
The number density of the variable dwarf stars is $\sim2\times10^3$ kpc$^{-3}$. 
Magnitude and color distributions of stars from the model predictions are 
consistent with those stars in the SXDF selected through our criteria 
and it might indicate that a few percent of the whole dwarf stars show optical bursts 
in our time samplings as shown in the bottom rows of Figure \ref{fig:colormaghist2}. 

Fractions of variable stars are one of interesting results. 
Fractions of variable stars as functions of colors of $R-i'$ and $B-V$ 
are nearly flat at a few percents as shown in Figure \ref{fig:colormaghist2}. 
\citet{tonry2005} investigated fractions of variable stars as a function 
of quartile variability down to variability of $0.015$ mag. 
They examined variability of stars in sequential 14-day observations 
down to $R\sim20$ mag at a superb photometric precision of $0.002$ mag. 
Their result indicates that a fraction of $0.016$ mag$/x$ of stars show variability 
above quartile variability of $x$ mag. 
Our detection threshold for variability of point sources corresponds to about $0.05-0.10$ mag 
and expected fractions are $\sim0.02-0.03$ using this equation. 
The fraction obtained by our survey for blue stars is $\sim0.03$ while that for red stars 
is $\sim0.05$ on average, and are slightly larger than the expected fraction. 
This discrepancy might be derived from some reasons described below. 
A young open cluster NGC2301, which \citet{tonry2005} targeted, is only 146 Myr old, 
where there should not be many variable populations. 
On the other hand, our pointings are towards the halo, 
where main components observed are old populations. 
Another reason can be the difference of time samplings; 
their observations are 14 consecutive days while our observations are sparse over years. 
Time scales which can be examined by each study are clearly different, 
which can cause these differences. 
FSVS studies recently investigated fractions of faint variable stars down to $V\sim24$ mag 
\citep*{moralesrueda2006,huber2006}. 
Their survey depths and directions are similar to ours, 
and their results for fractions of variable stars, $5-8\%$ \citep{huber2006} 
are also consistent with ours. 

\section{Summary}\label{sec:summary}

We investigated optical variability of faint objects down to $i'_{\rm{vari}}\sim25.5$ mag 
for variable components over $0.918$ deg$^2$ in the SXDF. 
Multi-epoch ($8-10$ epochs over $2-4$ years) imaging data obtained with Suprime-Cam on Subaru 8.2-m telescope
provided us the first statistical sample consisting of $1040$ optically variable objects by image subtraction 
for all the combinations of images at different epochs. 
We classified those variable objects into variable stars, SNe, and AGN, based on the optical morphologies, magnitudes, 
colors, optical-mid-infrared colors of the host objects, 
spatial offsets of the variable components from the host objects, and light curves. 
Although not all the variable objects were classified because of short time baselines of observations in the two fields, 
our classification is consistent with spectroscopic results and X-ray detections. 

We examined detection completeness for periodic variability and AGN variability. 
The completeness strongly depends on apparent magnitude. 
The completeness for AGN is $\sim1$ at $i'\sim21$ mag and deceases 
down to zero at $i'\sim24.5$ mag. 
Redshift dependence of the completeness calculated using light curves of PG quasar 
is small due to cancellation of time dilation and anti-correlation 
between rest-frame wavelength and variability. 
Among X-ray sources in the field, $36\pm2\%$ ($51\pm3\%$ 
for the bright sample with $i'<24.4$ mag) show optical variability. 
Variability detections of X-ray sources also show similar 
dependence on apparent magnitude to that from light curves. 

Number densities of variable objects for the whole sample, variable stars, 
SNe, and AGN as functions 
time interval $\Delta{t}$ and variable component magnitude $i'_{\rm{vari}}$ were obtained. 
About $5\%$ of all the objects show variability at $i'=21-22$ mag including host components 
although decreasing fractions towards fainter magnitude are caused 
by the detection limit for variable components. 
Number density of variable stars as a function of time interval $\Delta{t}$ 
is almost flat indicating that 
time scales of variability of these stars are short. 
Number density of SNe arrives at platau in time scales of a few months and 
that of AGN increases even in time scales of years. 
These results are consistent with expectations from typical time scales 
of their variability. 
Total number densities of variable stars, SNe, and AGN are 
$120$, $489$, and $579$ objects deg$^{-2}$, respectively. 

Variable stars show bimodal distributions in the color-magnitude diagrams. 
This indicates that these variable stars consist of blue bright ($V\sim22$ mag) 
variable stars of the halo population 
and red faint ($V\sim23.5$ mag) variable stars of the disk population. 
We selected a few candidates of RR Lyrae considering their large magnitude differences 
between maxima and minima and blue $B-V$ colors, 
The number density is $\sim10^{-2}$ kpc$^{-3}$ at a distance of $>150$ kpc from the Galactic center, 
which is consistent with extrapolations of previous results. 
Follow-up observations to determine the amplitudes and periods of variability might 
show that these candidates are at the outermost region of the Galactic halo. 

Our statistical sample of optically variable objects provides us unique opportunity for the studies 
such as AGN properties and SN rate, which will be topics in our following papers. 
There are planned large surveys such as 
Panoramic Survey Telescope and Rapid Response System \citep[Pan-Starrs;][]{kaiser2002}, 
Large Synoptic Survey Telescope \citep[LSST;][]{tyson2002}, 
and SuperNova Acceleration Probe \citep[SNAP;][]{aldering2002}. 
This work also provides basic information for such future wide and 
deep variability surveys. 

\acknowledgements
This work was supported in part with a scientific research grant (15204012) 
from the Ministry of Education, Science, Culture, and Sports of Japan (MEXT). 
M.A. is supported by a Grant-in-Aid for Young Scientists (B) from JSPS (18740118). 
This work is also supported in part with a scientific research grant (18072003) from the MEXT. 
We appreciate useful comments by Kimiaki Kawara and \v{Z}eljko Ivezi\'{c}. 
We are grateful to all members of the SXDS project. 
We also thank the anonymous referee for useful suggestions, which improve the manuscript. 

\begin{deluxetable}{ccccccc}
\tabletypesize{\scriptsize}
\tablewidth{0pt}
\tablecaption{Imaging Data Other than Optical Variability}
\tablehead{
\colhead{Wavelength} & 
\colhead{Telescope/Instrument} & 
\colhead{Band}&
\colhead{Detection Limit} & 
\colhead{Area\tablenotemark{a} [deg$^2$]} & 
\colhead{Variable\tablenotemark{b}} &
\colhead{Detection\tablenotemark{c}} 
}
\startdata
optical & Subaru/Suprime-Cam\tablenotemark{d} 
& $B$,$V$,$R_C$,$i'$,$z$ & $28.2$,$27.2$,$27.6$,$27.5$,$26.5$ mag\tablenotemark{e} & $0.918$ & $1040$ & $1040$ \\
X-ray & XMM-Newton/EPIC & 2.0-10.0 keV & $3\times10^{-15}$ erg s$^{-1}$ cm$^{-2}$ & $0.808$ & $936$ & $165$ \\
X-ray & XMM-Newton/EPIC & 0.5-2.0 keV & $1\times10^{-15}$ erg s$^{-1}$ cm$^{-2}$ & $0.808$ & $936$ & $91$ \\
mid-infrared & Spitzer/IRAC & 3.6$\mu$m-band & $22.0$ mag\tablenotemark{e} & $0.889$ & $1028$ & $995$ \\
\enddata
\tablenotetext{a}{Area overlapping with the variability survey field.} 
\tablenotetext{b}{Number of optically variable objects within the area.} 
\tablenotetext{c}{Number of optically variable objects detected in the bands within the area. }
\tablenotetext{d}{Derived from preliminary version of the SXDF catalogs in \citet{furusawa2007}.}
\tablenotetext{e}{Detection limits for optical and mid-infrared imaging are measured in AB magnitude.} 
\label{tab:surveyarea}
\end{deluxetable}
\begin{deluxetable}{crcrrcc}
\tabletypesize{\normalsize}
\tablewidth{0pt}
\tablecaption{Summary of Subaru Suprime-Cam $i'$-band Imaging Observations 
for Variability Detections}
\tablehead{
\colhead{field} & \colhead{epoch} & \colhead{date (UT)\tablenotemark{a}} & 
\colhead{$\Delta{t}$\tablenotemark{b}} & \colhead{$t_{\rm{exp}}$ [sec]} & 
\colhead{$\theta['']$\tablenotemark{c}} & \colhead{$m_{\rm{lim}}$\tablenotemark{d}}
}
\startdata
	SXDF-C & 1 & 02/09/29,30 & 0.0 & 2700 & 0.54 & 26.19\\
	SXDF-C & 2 & 02/11/01 & 32.6 & 1860 & 0.92 & 25.76\\
	SXDF-C & 3 & 02/11/02 & 33.5 & 1800 & 0.68 & 25.85\\
	SXDF-C & 4 & 02/11/05 & 36.7 & 2400 & 0.70 & 26.11\\
	SXDF-C & 5 & 02/11/09 & 40.5 & 2460 & 0.60 & 25.77\\
	SXDF-C & 6 & 02/11/27,29 & 59.8 & 4200 & 0.72 & 26.38\\
	SXDF-C & 7 & 02/12/07 & 68.4 & 3000 & 0.78 & 26.32\\
	SXDF-C & 8 & 03/10/20 & 385.7 & 5760 & 1.14 & 26.53\\
	SXDF-C & 9 & 03/10/21 & 386.5 & 7500 & 0.58 & 26.71\\
	SXDF-C & 10 & 05/09/28 & 1094.6 & 3600 & 1.00 & 26.04\\
\hline
	SXDF-N & 1 & 02/09/29,30 & 0.0 & 3300 & 0.56 & 26.26\\
	SXDF-N & 2 & 02/11/01 & 32.4 & 2640 & 0.96 & 25.88\\
	SXDF-N & 3 & 02/11/02 & 33.3 & 1800 & 0.68 & 25.86\\
	SXDF-N & 4 & 02/11/09 & 40.3 & 2100 & 0.64 & 25.78\\
	SXDF-N & 5 & 02/11/29 & 60.3 & 3300 & 0.74 & 26.27\\
	SXDF-N & 6 & 03/09/22 & 357.5 & 4264 & 0.60 & 26.37\\
	SXDF-N & 7 & 03/10/02 & 367.6 & 1500 & 0.70 & 25.88\\
	SXDF-N & 8 & 03/10/21 & 386.5 & 3000 & 0.72 & 26.14\\
\hline
	SXDF-S & 1 & 02/09/29,30 & 0.0 & 3000 & 0.52 & 26.28\\
	SXDF-S & 2 & 02/11/01 & 32.5 & 3600 & 1.04 & 25.91\\
	SXDF-S & 3 & 02/11/02 & 33.4 & 1800 & 0.70 & 25.83\\
	SXDF-S & 4 & 02/11/09 & 40.4 & 2580 & 0.66 & 25.60\\
	SXDF-S & 5 & 02/11/29 & 60.6 & 1500 & 0.82 & 26.00\\
	SXDF-S & 6 & 03/09/22 & 357.6 & 4500 & 0.54 & 26.45\\
	SXDF-S & 7 & 03/10/02 & 367.7 & 2040 & 0.68 & 26.00\\
	SXDF-S & 8 & 05/09/28 & 1094.6 & 3900 & 0.96 & 26.04\\
\hline
	SXDF-E & 1 & 02/09/29,30 & 0.0 & 3000 & 0.60 & 26.25\\
	SXDF-E & 2 & 02/11/01 & 32.5 & 3000 & 1.04 & 25.97\\
	SXDF-E & 3 & 02/11/02 & 33.4 & 1800 & 0.70 & 25.83\\
	SXDF-E & 4 & 02/11/09 & 40.4 & 2820 & 0.66 & 26.78\\
	SXDF-E & 5 & 02/11/29 & 60.6 & 1800 & 0.80 & 26.03\\
	SXDF-E & 6 & 02/12/07 & 68.6 & 1209 & 1.54 & 25.19\\
	SXDF-E & 7 & 03/09/22 & 357.6 & 6000 & 0.62 & 26.58\\
	SXDF-E & 8 & 03/10/02 & 367.7 & 1271 & 0.68 & 25.71\\
	SXDF-E & 9 & 03/10/21 & 386.7 & 1400 & 0.88 & 25.51\\
	SXDF-E & 10 & 05/09/28 & 1094.6 & 3600 & 0.96 & 26.11\\
\hline
	SXDF-W & 1 & 02/09/29,30 & 0.0 & 2400 & 0.54 & 26.14\\
	SXDF-W & 2 & 02/11/01 & 32.7 & 3000 & 0.96 & 25.96\\
	SXDF-W & 3 & 02/11/02 & 33.5 & 1800 & 0.66 & 25.84\\
	SXDF-W & 4 & 02/11/05 & 36.8 & 3060 & 0.76 & 25.98\\
	SXDF-W & 5 & 02/11/09 & 40.5 & 2100 & 0.64 & 25.97\\
	SXDF-W & 6 & 02/11/27,29 & 59.8 & 4200 & 0.74 & 26.39\\
	SXDF-W & 7 & 02/12/07 & 68.5 & 6483 & 1.04 & 26.34\\
	SXDF-W & 8 & 03/10/20,21 & 386.5 & 5460 & 0.66 & 26.46\\
\enddata
\tablenotetext{a}{Observed date in yy/mm/dd. When the images were stacked together 
with those at different dates, both dates are included.} 
\tablenotetext{b}{Days from the first observation in each field.} 
\tablenotetext{c}{FWHM of PSF in stacked images.} 
\tablenotetext{d}{\rm{L}imiting magnitude of $5\sigma$ in $2\farcs0$ aperture.} 
\label{tab:spcamobs}
\end{deluxetable}

\end{document}